\documentclass[aps,prl,twocolumn,superscriptaddress]{revtex4-1}
\usepackage{graphicx,bm,upgreek,bm,color,xcolor,mathtools,bbm,amssymb,amsmath,soul,titlesec}
\usepackage{tikz-feynman}
\usepackage{hyperref}

\def\onlinecite#1{\citenum{#1}}
\def\hH{{\hat{H}}}
\def\ha{{\hat{a}^{\phantom{\dagger}}}}
\def\had{{\hat{a}^\dagger}}
\def\hc{{\hat{c}^{\phantom{\dagger}}}}
\def\hcd{{\hat{c}^\dagger}}
\def\be{{\bf e}}
\def\bg{{\bf g}}
\def\bk{{\bf k}}
\def\bp{{\bf p}}
\def\bq{{\bf q}}
\def\bE{{\bf E}}
\def\br{{\bf r}}
\def\bRp{{{\bf R}_p}}
\def\bu{{\bf u}}
\def\a{{\alpha}}
\def\b{{\beta}}
\def\d{\delta}
\def\k{\kappa}
\def\s{{\sigma}}
\def\ve{\varepsilon}
\def\w{\omega}
\def\D{\partial}
\def\bs{{\boldsymbol\sigma}}
\def\<{\langle}
\def\>{\rangle}
\def\up{\uparrow}
\def\down{\downarrow}
\def\prefac{\frac{\hbar}{4m_{\rm e}^2 c^2}\,}
\def\Vtot{{\hat{V}_{\rm dRD}}}

\def\changes{}

\begin{document}

\title{Dynamic Rashba-Dresselhaus Effect}

\author{Martin Schlipf}
\affiliation{VASP Software GmbH, Sensengasse 8, 1090 Vienna, Austria}
\author{Feliciano Giustino}
\email{fgiustino@oden.utexas.edu}
\affiliation{Oden Institute for Computational Engineering and Sciences, The University of Texas at Austin, Austin, Texas 78712, USA}
\affiliation{Department of Physics, The University of Texas at Austin, Austin, Texas 78712, USA}

\date\today

\begin{abstract}
The Rashba-Dresselhaus effect is the splitting of doubly degenerate band extrema in semiconductors, accompanied by the emergence of counter-rotating spin textures and spin-momentum locking. Here we investigate how this effect is modified by lattice vibrations. We show that, in centrosymmetric non-magnetic crystals, for which a bulk Rashba-Dresselhaus effect is symmetry-forbidden, electron-phonon interactions can induce a phonon-assisted, dynamic Rashba-Dresselhaus spin splitting in the presence of an out-of-equilibrium phonon population.  In particular, we show how Rashba, Dresselhaus, or Weyl spin textures can selectively be established by driving coherent infrared-active phonons, and we perform \textit{ab initio} calculations to quantify this effect for halide perovskites.
\end{abstract}

\maketitle

\begin{figure}[b!]
\centering
\includegraphics[width=0.8\columnwidth]{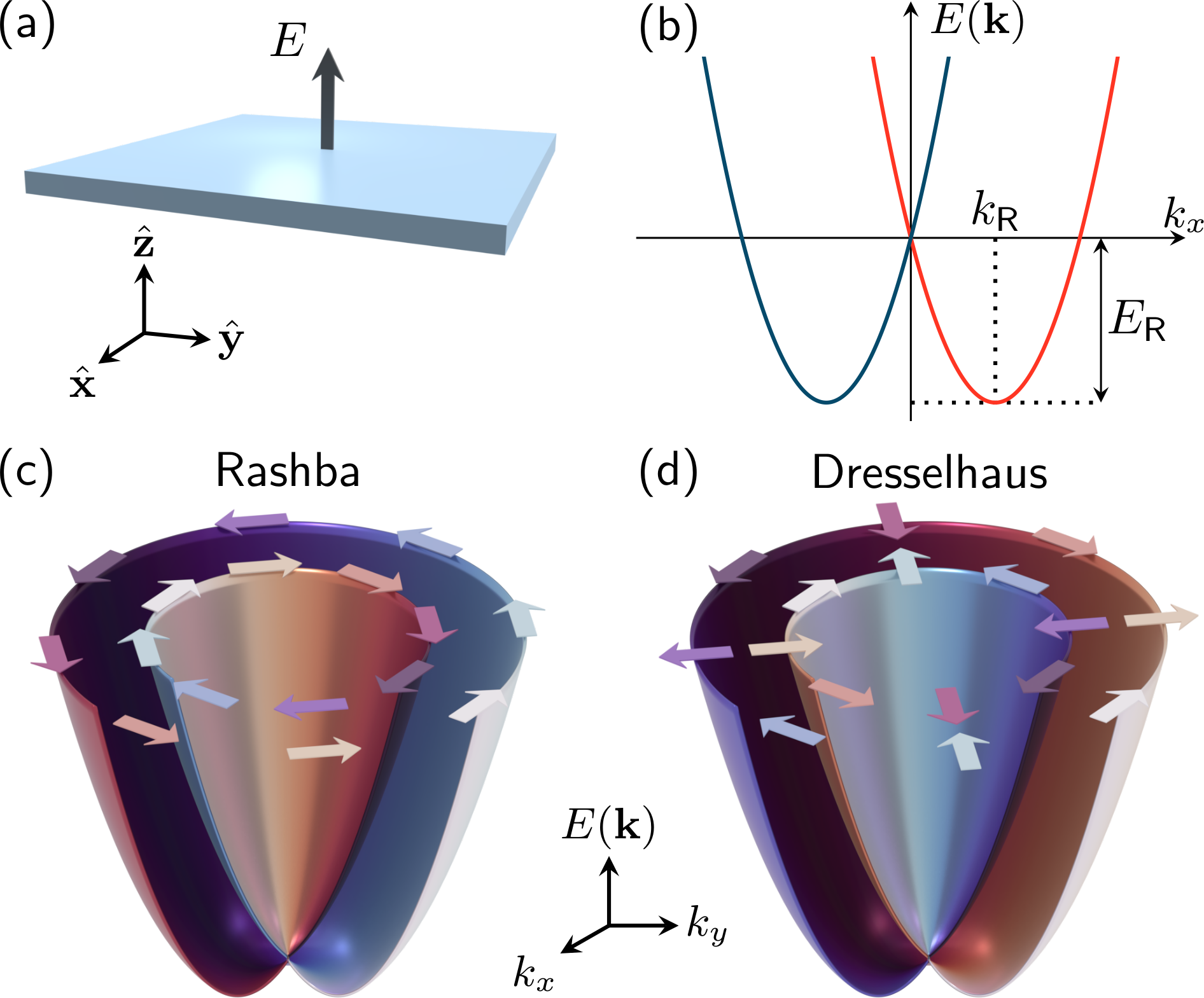}
\caption{
(a) In the Rashba-Dresselhaus effect the inversion symmetry of the crystal is broken, for example by an electric field. This lifts the spin degeneracy of the band minimum. (b) As a result, the band minimum is lowered by the Rashba energy $E_{\rm R}$, and is displaced by the Rashba wavevector $k_{\rm R}$. Crystal symmetry dictates whether the system assumes a (c) Rashba or a (d) Dresselhaus spin texture as indicated by the color of the bands and the arrows.}
\label{fig1}
\end{figure}

The interplay between crystal symmetry and spin-orbit coupling (SOC) is central to many recent advances in condensed matter physics. For example, SOC
{\changes
underpins the coupling between ferromagnetism and ferroelectricity in proper multiferroics~\cite{Kenzelmann2005,cm07}, the band inversion in topological insulators and quantum spin Hall insulators~\cite{Hasan2010}, the chiral properties of Weyl semimetals~\cite{Liu2014,Xu2015}, the anomalous Hall effect~\cite{Nagaosa2010}, and the emergence of skyrmions~\cite{Fert2017}.

An important manifestation of SOC is the Rashba-Dresselhaus (RD) effect, whereby degenerate electron bands split and acquire counter-rotating spin orientations (Fig.~\ref{fig1})~\cite{Rashba1959,dres55,brw15,Manchon2015}.  The RD effect can be used to generate and manipulate spin currents in Datta-Das transistors~\cite{Datta1990,Koo2009}, minimize spin dephasing in quantum devices~\cite{Zutic2004,Manchon2015}, and realize topological superconductivity and Majorana modes~\cite{Kezilebieke2020}. In current realizations, the RD effect is tuned electrically via a gate~\cite{Koo2009,Wunderlich2010} or chemically by balancing spin-orbit splitting and inversion-symmetry breaking~\cite{King2017}. Another theoretical possibility is to induce this effect by dynamically breaking inversion symmetry~\cite{emd16,Monserrat2017a,Niesner2018}.

In this work we develop the theory of dynamical control of the RD effect via phonon-assisted processes.  To enable maximum tunability, we focus on non-magnetic centrosymmetric crystals, for which the RD is forbidden. We identify the phonon symmetry selection rules that lead to RD spin splitting with besopke Rashba, Dresselhaus, or Weyl spin textures, and we identify an \textit{ab initio} descriptor to quantify this effect.  As an application of our approach, we calculate the dynamic RD effect for lead halide perovskites, and we show that it is within the detection range of current ultrafast experiments.
}

The RD effect can be understood starting from the standard spin-orbit coupling Hamiltonian:
 \begin{equation} \label{eq.1}
  V_{\rm SOC} = -\frac{e \,\hbar}{4m^2 c^2}\, \bE\cdot \bs \times \bp~,
 \end{equation}
where $\hbar$, $e$, $m$, $c$, $\bs$, and $\bp$ denote Planck constant, electron charge and mass, speed of light, Pauli vector, and electron momentum, respectively. $\bE$ is the electrostatic field experienced by the electrons. In the conventional Rashba effect one considers a uniform electric field along the Cartesian direction $\hat z$ as in Fig.~\ref{fig1}(a), and a parabolic electron band with minimum at the Brillouin-zone center, so that $V_{\rm SOC}$ is proportional to $\sigma_x k_y -\sigma_y k_x$, where $\bk$ is the Bloch wavevector of the electron. The related Dresselhaus coupling term is of the form {$\sigma_x k_x -\sigma_y k_y$}.  The resulting band structures and spin textures are illustrated in Fig.~\ref{fig1}(b) and (c)-(d), respectively.

In the absence of an external electric field, the vector $\bE$ in Eq.~\eqref{eq.1} is replaced by the gradient of the single-particle potential energy, $ \bE = (1/e)\nabla V$, which depends parametrically on the atomic coordinates, for example the Kohn-Sham potential. {\changes{}Here, we ask the following question:} under which conditions, in crystals with both space and time inversion symmetry, lattice vibrations can induce a dynamic RD effect via the dependence of $V$ on the atomic positions.

To answer this question, we consider a many-body Hamiltonian of interacting electrons and phonons, whereby atoms fluctuate around their equilibrium sites.
The phonon-induced spin-orbit coupling Hamiltonian in second-quantized notation reads:
 \begin{equation}\label{eq.2}
 \Vtot = N^{-\frac{1}{2}} {\sum}_{\,\bk, \bq,s,s'}\,\, g_{s' s} (\bk+\bq,\bk)\,
\hcd_{\bk+\bq s'}\hc_{\bk s}(\ha_{\bq}+\had_{-\bq})~,
 \end{equation}
where $\bk$ and $\bq$ are electron and phonon wavevectors belonging to a uniform Brillouin zone grid with $N$ points, $\hcd_{\bk s}$/$\hc_{\bk s}$ and $\had_{\bq}$/$\ha_{\bq}$ are creation/annihilation operators for electrons and phonons, respectively, and the subscript $s=1,2$ is the spinor label (`dRD' stands for `dynamic RD'). The spin-phonon coupling matrix elements are:
 \begin{equation} \label{eq.gb}
 g_{s' s}(\bk+\bq, \bk) = \prefac \< \psi_{\bk+\bq s'}| \bs \cdot \nabla (\Delta_{\bq} V)
\times \bp\, | \psi_{\bk s}\>~,
 \end{equation}
where $\psi_{\bk s}$ and $\psi_{\bk+\bq s'}$ are Pauli spinors, and $\Delta_{\bq} V$ represents the variation of the potential when the ions undergo a collective displacement along the phonon eigenmode with wavevector $\bq$. For notational simplicity we omit electron band and phonon branch indices; complete expressions are given in Supplementary Note~\ref{note-1}.

\textit{Thermal equilibrium --} For a system with both time-reversal and inversion symmetry, it is intuitive that phonon-induced spin splitting must be forbidden in thermal equilibrium, because thermal fluctuations do not break space and time symmetries. This is schematically illustrated in Supplementary Fig.~S\ref{fig-s1}(a)-(c) for the specific case of inversion symmetry.  Albeit intuitive, a rigorous proof of the lack of dynamic RD spin splitting in thermal equilibrium is nontrivial.
Here we only sketch the key steps of the proof, leaving the complete analysis to Supplementary Notes~\ref{note-2} and \ref{note-3}.

We consider the degenerate states $|\pm \bk, 1\>$ and $|\pm \bk, 2\>$, obtained from the ground state $|0\>$ by creating an extra electron in the conduction band, $|\bk s\> = \hcd_{\bk s}|0\>$.  The effect of $\Vtot$ on the energies of these states is analyzed using many-body perturbation theory to all orders.
Odd orders in perturbation theory involve the diagonalization of terms that contains products of the matrix elements $\<\bk s|\Vtot|\bk' s'\>$ with $s,s'=1,2$ and $\bk'=\pm \bk$. Each term of $\Vtot$ either creates or annihilates one phonon, therefore $\Vtot|\bk' s'\>$ and $|\bk s\>$ are orthogonal because they differ in their phonon occupations. As a result, odd-order terms in perturbation theory do not lead to spin splitting.
Even orders in perturbation theory involve products of spin-phonon matrix elements in the form:
 \begin{equation}\label{eq.4}
 p_{s' s} = \sum_{s_1}\sum_{s_2}\cdots\sum_{s_{2n-1}} g_{s' s_1} g_{s_1 s_2} \cdots g_{s_{2n-1} s}~,
 \end{equation}
where we consider the $2n$-th order and we omit the electron momenta for clarity.  In Supplementary Note~\ref{note-2} we show that parity and time-reversal symmetry require the spin-phonon matrix element to  transform as follows upon flipping the spins:
 \begin{equation}\label{eq.5}
 g_{\bar s' \bar s}(\bk',\bk) = (-1)^{s'-s+1}e^{i\varphi_{\bq}}\,g_{s' s}^*(\bk',\bk)~,
 \end{equation}
where  $\bar s$ is the spin label other than $s$, $\bq=\bk'-\bk$, and $\varphi_{\bq}$ is a phase associated with the transformation of vibrational eigenmodes upon inversion.  Using Eq.~\eqref{eq.5} inside Eq.~\eqref{eq.4} we find $p_{\bar s' \bar s}=(-1)^{s'-s} p^*_{s' s}$, hence $p_{12}^* = -p_{21}$. Since $\Vtot$ is Hermitian, we also have $p_{12}^*=p_{21}$. Therefore $p_{12}=p_{21}=0$ and even-order terms in perturbation theory do not lift the spin degeneracy.  For completeness in Supplementary Note~\ref{note-3} and \ref{note-4} we provide the expression for the second-order and fourth-order energy shifts resulting from $\Vtot$, and in Supplementary Note~\ref{note-5} we generalize the above reasoning to finite temperature. This proof confirms the lack of a phonon-induced spin splitting in thermal equilibrium.

\textit{Out-of-equilibrium phonons --} Next we {\changes{}move to} a non-thermal phonon population. We consider a quantum state that is not an eigenstate of the unperturbed Hamiltonian.
As a representative example we focus on coherent Glauber states, {\changes{}which are minimum-uncertainty wavepackets with} a probability distribution similar to the ground state, but rigidly translated away from the equilibrium coordinate, as sketched in Supplementary Fig.~S\ref{fig-s1}(d).  These states are of particular interest since they can be generated and detected using ultrafast pump-probe techniques~\cite{Salen2019}.

A coherent state associated with the phonon of momentum $\bq$ and frequency $\w_\bq$ can be written as:
 \begin{equation}\label{eq.8}
 |\bk s, u\> = \hcd_{\bk s} \,{\rm exp}(-N u^2/2) \,{\rm exp}\big( N^{1/2} u \,\had_\bq \,\big) |0\>~,
 \end{equation}
where $u(t)$ is the instantaneous fluctuation amplitude (this expression is for $\bq=0$; when $\bq\ne 0$ the exponent is replaced by $\had_\bq+\had_{-\bq}$, see Supplementary Note~\ref{note-6}).  The expectation value of the unperturbed Hamiltonian over this coherent state is $E_{\bk ,u} = \ve_{\bk} + u^2 N\,\hbar\w_{\bq}$, where $\ve_{\bk}$ and $\w_{\bq}$ are the non-interacting electron energy and phonon frequency, respectively. Therefore $u^2$ quantifies the number of excited phonons per crystalline unit cell. Since $|\bk s, u\>$ is an eigenvector of the annihilation operator $\had_\bq$, the expectation value of the atomic displacements $\Delta {\bm \tau}_\kappa$ on this state is nonzero:
 \begin{equation}\label{eq.10}
 \< \Delta {\bm \tau}_\kappa \>_t =  u(t)\,2 (\hbar/2M_\kappa
\w_\bq)^{1/2}\,\be_{\kappa}~,
 \end{equation}
where $M_\kappa$ is the mass of the $\kappa$-th atom, and $\be_{\kappa}$ is the phonon polarization. Unlike vibrations in thermal equilibrium, this state induces time-dependent inversion symmetry breaking, hence spin splitting should be allowed. It remains to be seen whether the energy splitting and the accompanying spin texture bear any resemblance with the conventional RD effect.

To probe the consequences of inversion symmetry breaking, we evaluate the energy expectation value of the total Hamiltonian:
  \begin{equation}\label{eq.expt.values.2b}
  \<\bk s, u|\,\hH_0+\Vtot\,|\bk s', u\> =\d_{s's}E_{\bk,u} + \,2\, \d_{\bq,0} \,u \, g_{s s'}(\bk,\bk)~.
  \end{equation}
From the Kronecker delta $\d_{\bq,0}$ we see that spin splitting is only allowed when $\bq\!=\!0$. This selection rule arises from the fact that only zone-center phonons can break the inversion symmetry of the crystal. We note that this result was derived for idealized infinite crystals; in the realistic case of a finite crystal of size $L$, the selection rule only requires $|\bq|\le \pi/L$. We proceed to consider $\bq\!=\!0$ coherent states.

Time-reversal invariance implies that the vibrational eigenmodes can be chosen to satisfy $\be_{\k}(-\bq) = \be^*_{\k}(\bq)$. Furthermore, invariance under parity requires $\be_{\tilde \kappa}(-\bq) = -e^{i\varphi_{\bq}}\be_{\kappa}(\bq)$, where the ion $\tilde \k$ is the inversion partner of $\k$, and $\varphi_{\bq}$ is the same as in Eq.~\eqref{eq.5} \cite{Maradudin1968}.  For zone-center phonons, these relations imply that the eigenmodes have definite parity and $\varphi_\bq=\pi\,/\,0$ for even/odd phonons.  Using this observation in Eq.~\eqref{eq.5}, we obtain $g_{s \bar s}(\bk,\bk) = \mp\, g_{s \bar s}(\bk,\bk)$ for even/odd phonons. It follows that spin splitting is allowed only for zone-center odd-parity modes.

To elucidate the nature of the spin-split bands, we diagonalize Eq.~\eqref{eq.expt.values.2b} by considering a parabolic band of mass $m^*$ with extremum at the zone center. After performing a Taylor expansion of $g_{s s'}(\bk,\bk)$ at small $\bk$, and noting that $g_{s s'}(0,0)=0$ as a consequence of inversion symmetry (see Supplementary Note~\ref{note-7}), we obtain:
{\changes
 \begin{equation}\label{eq.split.dftb}
 \Delta \ve_{\bk} = \pm 2 E_{\rm R}\, |\bk|/k_{\rm R}~,
 \end{equation}
where the Rashba energy and wavevectors are $E_{\rm R} = \hbar^2 k_{\rm R}^2/2 m^*$ and $ k_{\rm R} =  \lVert \hat \bk \cdot \nabla_\bk\, g_{s s'}(\bk, \bk) \rVert u \sqrt{2}m^*/\hbar^2$, respectively}. Here $\hat \bk = \bk/|\bk|$ and $\lVert \cdot \rVert$ is the Frobenius norm in the spin indices.  Eq.~\eqref{eq.split.dftb} shows that the splitting vanishes at $\bk=0$, and increases linearly with $\bk$ away from the zone center.  This is precisely the hallmark of the Rashba effect encoded in Eq.~\eqref{eq.1}. The resulting electronic bands are sketched in Fig.~\ref{fig1}(c). In the case of driven oscillations, the splitting and the associated spin texture will fluctuate with the amplitude $u(t)$.

In Supplementary Note~\ref{note-8} we estimate the Rashba energy in Eq.~\eqref{eq.split.dftb} for the Fr\"ohlich interaction associated with longitudinal-optical (LO) phonons in polar crystals~\cite{Froehlich1954}. Counter-intuitively, we find that the Fr\"ohlich coupling induces a negligible RD splitting, because the characteristic singularity of the polar coupling at long-wavelength is eliminated when taking the gradient $\nabla (\Delta V_{\bq\nu})$ in Eq.~\eqref{eq.gb}. Combined with the fact that only transverse-optical (TO) modes are excited by light in bulk crystals, this result suggests that the search for a dynamic RD effect should focus on zone-center TO modes of ungerade (u) symmetry, namely the IR-active modes of polar crystals. Since in centrosymmetric crystals the modes cannot simultaneously be IR- and Raman- active, we must rule out the possibility of directly realizing RD splitting using a Raman-active mode (unless such a mode is used to excite an IR mode via nonlinear couplings \cite{Radaelli2018}).

{\changes Now we proceed to examine the spin texture associated with the above RD splitting}. To this end we consider the symmetry of the spin-phonon matrix element in Eq.~\eqref{eq.gb}.  Let us call $\bg$ the $2\times 2$ matrix with elements $g_{ss'}(\bk,\bk)$ in the spinor labels $s,s'$.  Using $\bk\cdot \bp$ perturbation theory and the algebra of Pauli matrices, in Supplementary Note~\ref{note-9} we show that $\bg$ can be expressed as:
  \begin{equation}\label{eq.Gmat}
  \bg =  \sum_{\a\b}k_\a G_{\a\b }\,\s_\b~,
  \end{equation}
where the components of the real-valued $3\times 3$ matrix $G_{\a\b}$ take the {\changes{}form
$ G_{11} = (\hbar^2/8 m_e^2 c^2) \< 0, 1|[ \bs\times\nabla]_\a \Delta_{0} V ) | 0,2\> + {\rm c.c.}~, $}
and similarly for the other elements. The complete matrix is given in Supplementary Note~\ref{note-9}, Eq.~(S81). Direct inspection of Eq.~\eqref{eq.Gmat} shows that the isotropic part of $G_{\a\b}$ leads to couplings like {$k_x\s_x + k_y\s_y + k_z \s_z$}, and therefore it induces a Weyl spin texture.  The traceless symmetric part of $G_{\a\b}$ leads to couplings like {$k_x \s_x - k_y \s_y$ or} $k_x\s_y+k_y\s_x$, hence it induces a Dresselhaus spin texture.  The antisymmetric part of $G_{\a\b}$ leads to couplings like $k_x\s_y-k_y\s_x$, which correspond to a Rashba spin texture.  Altogether, the present findings indicate that a dynamic RD effect with similar phenomenology as the conventional (static) effect is theoretically possible, and that the spin texture can be tuned by exciting IR modes of select symmetry.

\textit{Proposed experiments --} To realize a dynamic RD effect using out-of-equilibrium coherent phonons, we propose to perform THz pump/optical probe experiments on halide perovskites, {\changes{}following recent work on methylammonium lead iodide (MAPI, CH$_3$NH$_3$PbI$_3$)~\cite{Niesner2018,Liu2020, Frohna2018} which is known to have a large SOC~\cite{Even2013}}.
\begin{figure}
\centering
\includegraphics[width=0.9\columnwidth]{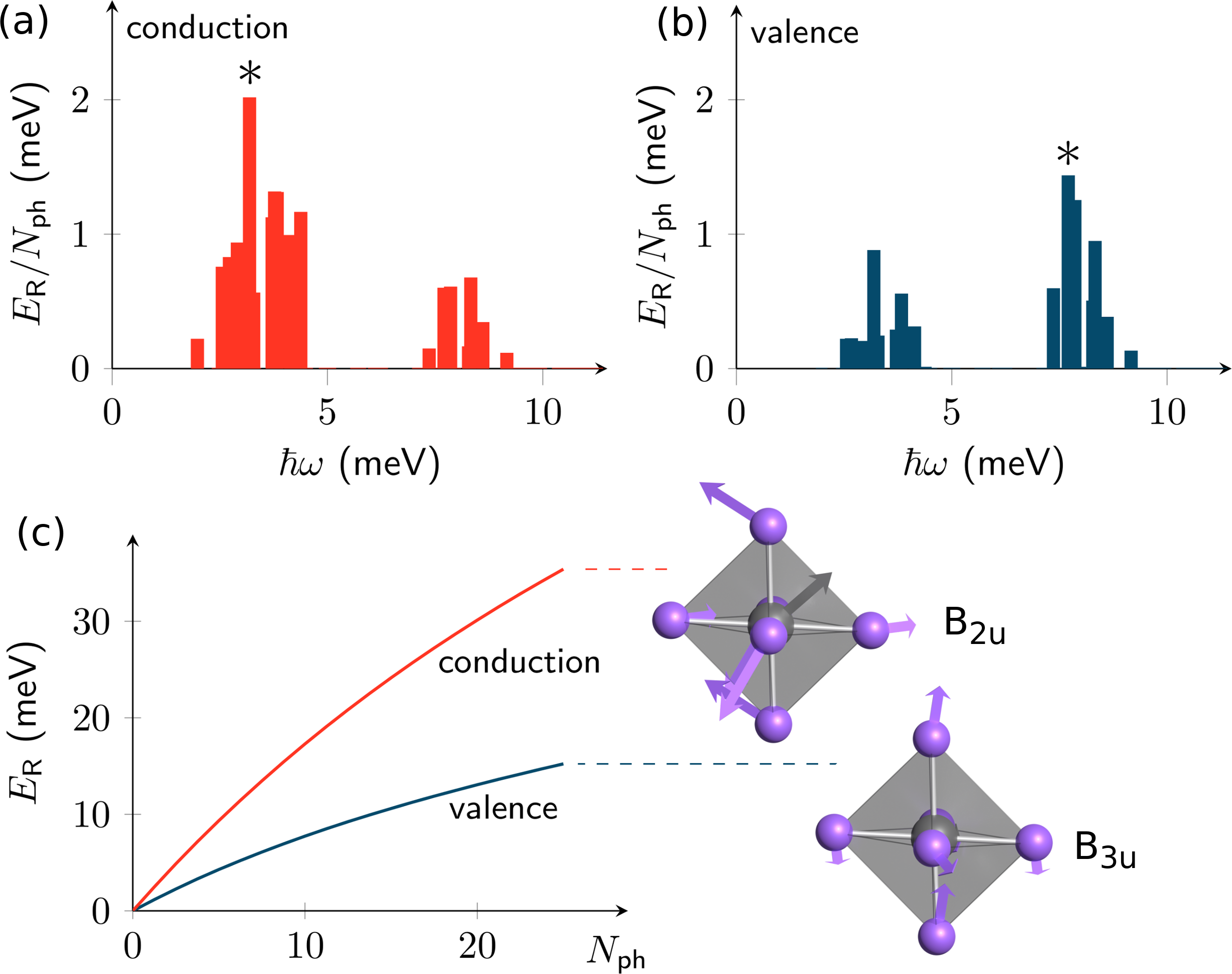}
\caption{
(a) Rashba energy for every IR-active mode in MAPI and for the conduction band bottom.  The energy is evaluated for $N_{\rm ph}=u^2 = 1$ excited phonon per orthorhombic unit cell.  (b) Same as in (a), but for the valence band of MAPI.  (c) Dependence of the Rashba energy on the number of phonons per unit cell excited in the modes marked by asteriscs in (a) and (b): the $B_{2{\rm u}}$ Pb-I-Pb rocking mode at 3.2~meV, and the $B_{3{\rm u}}$ Pb-I stretching mode at 7.7~meV.  These modes are schematically illustrated by the ball-and-stick models.
}
\label{fig2}
\end{figure}
Below 160\,K MAPI crystallizes in a orthorhombic structure with centrosymmetric space group $Pnma$ (Supplementary Fig.~S\ref{fig-s4}). This system admits 20 IR-active optical phonons, associated with the deformation of the PbI$_6$ octahedra~\cite{Perez-Osorio2015}. Figure~\ref{fig2} shows the Rashba energy $E_{\rm R}$ for each of these modes, as calculated using Eq.~\eqref{eq.split.dftb} (the calculation details are provided in the Supplementary Methods). For the conduction band, we find that the $B_{2{\rm u}}$ mode at 3.2~meV, which corresponds to the Pb-I-Pb rocking vibration, provides the largest spin splitting [Fig.~\ref{fig2}(a)].  For the valence bands, the largest splitting is found with the $B_{3\rm u}$ mode at 7.7~meV, corresponding to the Pb-I stretching vibration [Fig.~\ref{fig2}(b)].
\begin{figure*}
\centering
\includegraphics[width=0.8\textwidth]{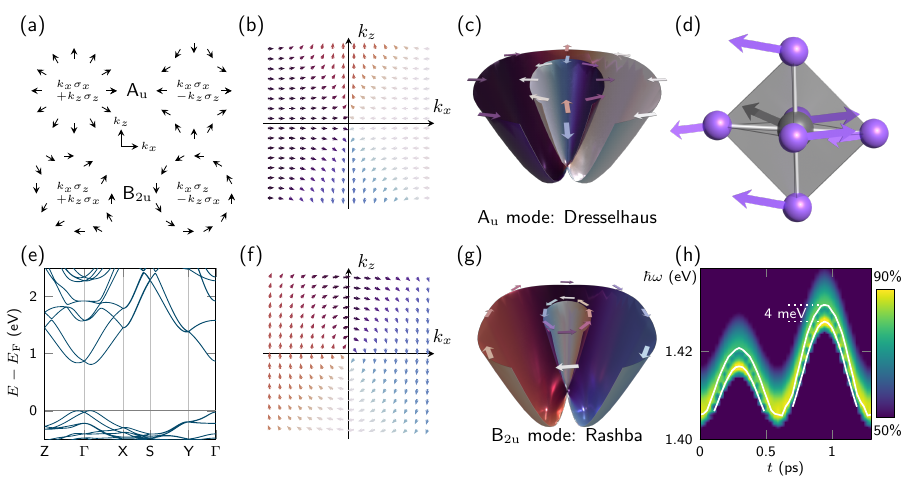}
\caption{
{\changes
(a) Spin-split band structure resulting from coherently driving the $B_{2{\rm u}}$ mode in MAPI.  (b)-(c) Upon driving the $A_{{\rm u}}$ mode, the spin texture of the conduction band assumes a Dresselhaus pattern, similar to Fig~\ref{fig1}(d). The ball-and-stick model of this mode is shown in (d).  (e) Allowed spin textures in the $k_x$-$k_z$ plane for $A_{\rm u}$ and $B_{2{\rm u}}$ modes. The $A_{\rm u}$ mode is compatible with a Weyl or Dresselhaus texture, the $B_{\rm 2u}$ mode admits a Dresselhaus or Rashba texture.  (f)-(g) Driving the $B_{2{\rm u}}$ mode results into a Rashba spin texture, similar to Fig~\ref{fig1}(c).  (b) and (f) show the spin texture for the conduction band bottom projected on the $k_x$-$k_z$ plane.  (c) and (g) show the spin texture color-coded on the band structure.  (h) The RD band splitting associated with a periodically driven B$_{2{\rm u}}$ mode induces a fluctuating peak-dip-hump structure in the PL spectrum of MAPI.}
}
\label{fig3}
\end{figure*}

The maximum RD splitting achievable in experiments is limited by the stability of the crystal under illumination. According to Lindemann's criterion~\cite{Lindemann1910}, melting occurs when the ionic displacements exceed approximately 10\% of the equilibrium bond length.  An upper bound to the displacement in Eq.~\eqref{eq.10} is $\Delta \tau_{\rm max} = 2 ( N_{\rm ph}\hbar/2M_{\rm min}\w_\bq)^{1/2}$, where $M_{\rm min}$ is the smallest atomic mass. Using the mass of iodine and the frequency of the $B_{2{\rm u}}$ mode, and setting $\Delta \tau_{\rm max}$ to 10\% of the Pb-I bond length (3.2~\AA), we find $N_{\rm ph} = 50$.  Figure~\ref{fig2}(c) shows that, even for considerably smaller vibrational amplitudes, Rashba energies in excess of $30$~meV should be within reach.

{\changes
A representative band structure snapshot corresponding to the $B_{2{\rm u}}$ mode is shown in Fig.~\ref{fig3}(a). These bands describe the system at a given time $t$. Since the $B_{2{\rm u}}$ mode oscillates with a period of $1.3$~ps, and the electron lifetime near the conduction band edges of MAPI is $\sim$10~fs at 160~K~\cite{Schlipf2018}, electrons  effectively experience a quasi-static potential. Therefore the notion of `instantaneous' bands is meaningful in this case.

The spin textures for the $A_{\rm u}$ and $B_{2{\rm u}}$ excitations are shown in Fig.~\ref{fig3}(b), (c), (f),
and (g).} To understand the spin patterns, let us consider the elements of the matrix $G$ in Eq.~\eqref{eq.Gmat} for odd-parity modes in the $D_{2\rm h}$ point group.  Due to the presence of spin-orbit coupling, we must consider the double group to correctly represent the spinors.  If $\Gamma_{\a}$ is the irreducible representation of the operator $[\bs \times \nabla]_\a (\Delta_{0} V)$ appearing in Eq.~\eqref{eq.Gmat}, group theory~\cite{Koster1963} determines which matrix elements of $G_{\a\b}$ can be nonzero.  Using Supplementary Table~S\ref{tab-s1} A, we find that for any given representation $\Gamma_{\a}$ only a single component $\b$ yields a nonzero element.  In particular, these are $\b = 3$ ($\Gamma_{\a} = B_{\rm1g}$), $\b = 2$ ($\Gamma_{\a} = B_{\rm2g}$), and $\b = 1$ ($\Gamma_{\a} = B_{\rm3g}$), which is exactly the representation of $\s_\b$ as expected.  Next, we need to determine $\Gamma_{\a}$ for a given phonon mode. In Supplementary Table~S\ref{tab-s1} B we give the product of the irreducible representations of Pauli matrix, gradient, and phonon mode for the different Cartesian directions, and we indicate the resulting allowed couplings between $\bk$ and $\bs$, corresponding to nonzero elements of the matrix $G_{\a\b}$.  This analysis indicates that the allowed couplings correspond to $k_x \s_x$, $k_y \s_y$, $k_z \s_z$ for the $A_{\rm u}$ mode, $k_x \s_y$, $k_y \s_x$ for the $B_{1\rm u}$ mode with polarization along $z$, $k_x \s_z$, $k_z \s_x$ for the $B_{2\rm u}$ mode with polarization along $y$, and $k_y \s_z$, $k_z \s_y$ for the $B_{3\rm u}$ mode with polarization along $x$. The orthorhombic symmetry does not dictate the sign relationships between the allowed couplings, so the signs must be {determined} by direct inspection of the spin texture. In Fig.~\ref{fig3}(e) we show that a $B_{\rm 2u}$ mode is compatible with a Rashba or Dresselhaus spin pattern and the $A_{\rm u}$ mode with a Weyl or Dresselhaus texture. By manually inspecting the spin patterns we confirm that our \textit{ab initio} calculations follow the above symmetry considerations: For the $A_{{\rm u}}$ mode a Dresselhaus pattern is established [Fig.~\ref{fig3}(b) and (c)], whereas the $B_{2{\rm u}}$ mode leads to a Rashba pattern [Fig.~\ref{fig3}(f) and (g)].

The key experimental challenge to realize the dynamic RD effect proposed here, is to pump coherent oscillations in the low-THz range (the $B_{2{\rm u}}$ mode of MAPI has a frequency $\sim$0.8~THz) without damaging the sample~\cite{Udina2019,Huber2015}.  A promising approach that has emerged during the past few years is to exploit resonant sum-frequency excitation processes~\cite{Salen2019,Udina2019}.  This approach was demonstrated for Te crystals (2~THz source)~\cite{Huber2015}, SrTiO$_3$ (3~THz)~\cite{Kozina2017}, Bi$_2$Se$_3$ (1~THz)~\cite{Melnikov2018}, LiNbO$_3$ (1~THz, 4~THz, and 18~THz)~\cite{Hirori2011, Dastrup2017,Cavalleri2018}, and CdWO$_4$ (2~THz)~\cite{Johnson2019}. More recently, single-cycle THz pulses have been used to drive coherent oscillations in MAPI~\cite{Liu2020}. {\changes{}In this work, the authors were able to
detect coherent oscillations in the differential reflectivity spectra lasting
as long as 5~ps, which should be more than sufficient to realize the effect proposed here.}

{\changes
A possible way to probe the driven dynamic RD effect would be via time-resolved ARPES experiments.
In such experiments the shape of the constant-energy cuts should oscillate between a circle in absence of RD splitting, as observed in Ref.~\onlinecite{Puppin2020}, and two distinct circles for Rashba-split bands.
Alternatively one could monitor optical properties such as the oscillator strength of the first excitonic peak~\cite{Davies2018}, which is expected to be sensitive to the fluctuating RD splitting of the band extrema~\cite{Liu2020}. As an illustrative example, we have calculated the time-revolved photoluminescence (PL) spectrum of MAPI with the $B_{2{\rm u}}$ coherently driven (see Supplementary Methods). Fig.~\ref{fig3}(h) shows that the dynamic RD effect induces a discernible fluctuating peak-dip-hump structure in the PL signal.
}

In summary, we developed an {\it ab initio} theory of the dynamic, phonon-assisted Rashba-Dresselhaus effect. We showed that, in centrosymmetric non-magnetic crystals, this effect can be realized and controlled by driving coherent oscillations of infrared-active optical modes, and we established the relation between the mode symmetry and the appearence of a Rashba, Dresselhaus, of Weyl spin texture. We made quantitative predictions for lead-based halide perovskites and proposed possible experimental realizations. Our work shows that {\changes{}phonons may provide new strategies for harnessing the spin degrees of freedom in quantum materials}.

\vspace{3pt}
\begin{acknowledgments}
Work by M.S. was supported by the Leverhulme Trust under award RL-2012-001. F.G. was supported by the Computational Materials Sciences Program funded by the U.S. Department of Energy, Office of Science, Basic Energy Sciences, under Award DE-SC0020129. {\changes{}The authors acknowledge the Texas Advanced Computing Center (TACC) at The University of Texas at Austin for providing HPC resources via the Frontera LRAC project DMR21002.}

\end{acknowledgments}




\def\onlinecite#1{\citenum{#1}}
\renewcommand\theequation{S\arabic{equation}}
\renewcommand{\thesection}{\arabic{section}}
\renewcommand{\thetable}{\arabic{table}}
\titleformat{\section}{\normalfont\bfseries}{\thesection.}{5pt}{}
\renewcommand{\baselinestretch}{1.15}
\setlength{\parindent}{0pt}
\setlength{\parskip}{5pt}
\titlespacing{\section}{0pt}{6pt}{0pt}

\def\hH{{\hat{H}}}
\def\ha{{\hat{a}^{\phantom{\dagger}}}}
\def\had{{\hat{a}^\dagger}}
\def\hc{{\hat{c}^{\phantom{\dagger}}}}
\def\hcd{{\hat{c}^\dagger}}
\def\bA{{\bf A}}
\def\be{{\bf e}}
\def\bg{{\bf g}}
\def\bk{{\bf k}}
\def\bp{{\bf p}}
\def\bq{{\bf q}}
\def\br{{\bf r}}
\def\bRp{{{\bf R}_p}}
\def\bu{{\bf u}}
\def\a{{\alpha}}
\def\b{{\beta}}
\def\g{{\gamma}}
\def\d{\delta}
\def\k{{\kappa}}
\def\s{{\sigma}}
\def\l{{\lambda}}
\def\ve{\varepsilon}
\def\w{\omega}
\def\D{\partial}
\def\bs{{\boldsymbol\sigma}}
\def\<{\langle}
\def\>{\rangle}
\def\up{\uparrow}
\def\down{\downarrow}
\def\prefac{\frac{\hbar}{4m_{\rm e}^2 c^2}\,}
\def\prefact{\frac{\hbar^2}{4m_{\rm e}^2 c^2}\,}
\def\Vtot{{\hat{V}_{\rm dRD}}}
\def\Ecvk{E_{{\rm cv}\bk}}

\tikzfeynmanset{compat=1.0.0}
\renewcommand{\figurename}{Fig.~S\!\!}
\renewcommand{\tablename}{Table~S\!\!}





\bibliography{biblio}{}

\clearpage
\newpage
\widetext

\setcounter{equation}{0}
\setcounter{figure}{0}
\setcounter{secnumdepth}{1}

{\large \textbf{Supplementary Materials}}

\section*{Supplementary Methods}
To determine the Rashba energy and the spin texture of methylammonium lead iodide we employ density-functional theory (DFT).  We relax the position of the atoms starting from the experimental $Pnma$ structure.\cite{Baikie2013} We evaluate the dynamical matrix using density-functional perturbation theory,\cite{Baroni2001} and from there we obtain phonon frequencies and vibrational eigenmodes.  The mode assignment is the same as in Ref.~\onlinecite{Perez-Osorio2015}.  The Rashba energies in Fig.~2 of the main manuscript are then obtained by displacing the atoms along these vibrational eigenmodes (see Supplementary Note~\ref{note-7}) with the amplitude $u^2$. The expectation value of the Pauli matrices with the Kohn-Sham wavefunctions yields the spin texture and allows to distinguish between a Rashba and a Dresselhaus spin splitting.

For our calculations we utilize the {\scshape Quantum Espresso} software.\cite{QE2017} We use ultrasoft pseudopotentials\cite{Vanderbilt1990} including the $d$ semicore states of Pb and I. We approximate the exchange-correlation functional with PBE\cite{PBE} and include plane waves up to 40~Ry and 200~Ry for wave functions and density, respectively.  We sample the Brillouin zone with a $4 \times 4 \times 4$ $\Gamma$-centered $\bk$-point grid to obtain the self-consistent density. We perform a non-self-consistent calculation with a dense sampling along the high-symmetry lines (spacing 0.0014~\AA$^{-1}$) to determine the Rashba energy, and with a dense $16 \times 16$ grid in a square (width 0.107~\AA$^{-1}$) surrounding the $\Gamma$ point to obtain the spin texture. The orientation of the square is perpendicular to the symmetry of the mode, e.g., for the $B_{2\rm u}$ mode the square is in the $k_x$-$k_z$ plane.

For the analysis of the symmetry, we use the Isotropy software.\cite{Isotropy} We analyze the invariants of the space group 62 for the even irreducible representation corresponding to the spin component coupled to two odd irreducible representations for $\bk$ point and phonon mode. We focus in particular on all trilinear invariants that couple all of these representations.

The effect of a coherent phonon on the electronic structure is approximated
by a frozen-phonon calculation. This approximation is meaningful because the frozen phonon method
provides the first term in the series expansion of the electron eigenvalues over the atomic
displacements. By way of example, let us
consider the simplest case of a one-dimensional
oscillator of mass $m$ and frequency $\w$. The wavefunction of a coherent phonon $|u\>$ can be written
in terms of the normal mode coordinate $x$ as:
  \begin{equation}
  \<x|u\> = \left(\frac{m\w}{\pi\hbar}\right)^{1/4}
        \exp\left[-\frac{m\w}{2\hbar}\left(x-\sqrt{\frac{2\hbar}{m\w}} u\right)^2\right].
  \end{equation}
Using this expression, the expectation value of the electronic Hamiltonian $\hH$ as a parametric operator
of the normal coordinate $x$ can be written as:
  \begin{equation}
  \<u|\hH|u\> =
     \hH(x=2u\sigma)
        + \frac{\sigma^2}{2}\!\left.\frac{\D^2 \hH}{\D x^2}\right|_{x=2u\sigma}
         \!\!\!+\,\, \mathcal{O}(\sigma^4),
  \end{equation}
where $\sigma = \sqrt{\hbar/2m \w}$  is the zero-point
vibrational amplitude. This relation shows that the expectation value of the electronic
Hamiltonian over a coherent phonon, $\<u|\hH|u\>$, can be evaluated using
a frozen phonon calculation with displacements $x=2u\sigma$, with an error that is of
second order in the zero-point amplitude. The neglected second-order term
on the r.h.s. of the above equation corresponds
to the Debye-Waller self-energy correction,\cite{Giustino2017}
which yields a spin-independent shift of the bands.

To analyze the photo-luminescence spectrum ${\rm PL}(E)$, we {assume constant transition matrix elements and low photoexcitation densities so that the Fermi-Dirac function can be replaced by a decaying potential. Then the PL spectrum can be expressed as\cite{YuCardona}
\begin{equation}
    {\rm PL}(E) \propto \sum_{{\rm cv}\bk} \exp\Bigl[-\frac{E - E_{\rm g}}{kT} \Bigr] \delta(E - \Ecvk)~.
\end{equation}
Here, $E_{\rm g}$ is the band gap, $\Ecvk = E_{{\rm c}\bk} - E_{{\rm v}\bk}$, and we used a temperature $T = 100$~K.}
Convergence of the PL spectrum requires an ultra-dense $200 \times 140 \times 205$ mesh to perform the $\bk$-point summation. {Including all points is not computationally feasible, therefore} we limit ourselves to $\bk$-points within a sphere of $0.08 \cdot 2\pi/a$ surrounding the $\Gamma$ point. Additionally using the symmetry of the crystal allows us to further reduce the sphere to a single octant leaving us with a total of 2443 $\bk$-points. To avoid artifacts caused by {smearing} methods, we construct a Delaunay triangulation. We use a linear tetrahedron method on the result to obtain the {joint} density of states.

We evaluate ${\rm PL}(E)$ for 25 different time-points displacing the atoms according to the $B_{\rm 2u}$ mode with the largest Rashba energy. The period of the oscillation is 1.3~ps corresponding to a phonon energy of 3.2~meV. The amplitude of the mode is chosen such that the maximum distortion {of Pb and I}
{corresponds to a displacement of {0.06}~\AA\ from equilibrium ($<\!2\%$ of the Pb-I bond length).}
We normalize ${\rm PL}(E)$ to the largest value obtained for any structure.

\section*{Supplementary Notes}

\section{Spin-phonon coupling arising from spin-orbit interactions}\label{note-1}

Here we introduce the notation employed throughout the manuscript to analyze the spin-phonon coupling arising from spin-orbit interactions.  The spin-orbit coupling (SOC) Hamiltonian is given by:\cite{Schiff}
  \begin{equation} \label{eq.soc}
  V_{\rm SOC}\big(\br;\{\tau_{\k\a p}\}\big) = \prefac \bs\cdot
  \nabla V\big(\br;\{\tau_{\k\a p}\}\big)\times \bp~,
  \end{equation}
where $\hbar$ is the Planck constant, $m_{\rm e}$ the electron mass, and $c$ the speed of light. The Pauli vector is given by $\bs = \sum_{\a=1}^3\s_\a \bu_\a$, where $\bu_\a$ are the Cartesian unit vectors and $\sigma_\a$ are the Pauli matrices. $V\big(\br; \{\tau_{\k\a p}\}\big)$ is the mean-field potential experienced by the electrons, such as for example the Kohn-Sham potential, which depends parametrically on the ionic coordinates $\{\tau_{\k\a p}\}$. Here $\k$ denotes the ion in the unit cell, $p$ denotes one of the $N$ unit cell contained in the Born-von K\'arm\'an (BvK) supercell, and $\bp$ is the electron momentum operator. In second-quantized notation, the SOC Hamiltonian in Eq.~\eqref{eq.soc} takes the form:
  \begin{equation} \label{eq.soc.sq}
  \hat{V}_{\rm SOC} = \sum_{n\bk s,n'\bk's'} \< \psi_{n\bk s}| V_{\rm SOC}\big(\br;
  \{\tau_{\k\a p}\}\big) | \psi_{n'\bk' s'}\> \hcd_{n\bk s}\hc_{n'\bk's'}~,
  \end{equation}
where $\psi_{n\bk s}(\br)$ denotes a two-component Pauli spinor, $\psi_{n\bk s}(\br) = \left[ \psi_{n\bk s}^{(1)}(\br) \,\, \psi_{n\bk s}^{(2)}(\br) \right]^{\rm T}$, which is solution of the Hamiltonian at clamped ions for the eigenvalue $\ve_{n\bk s}$. The $\br$-integral is performed over the BvK supercell, and the operators $\hcd_{n\bk s}$/$\hc_{n\bk s}$ create/annihilate an electron in the spinor state $n\bk s$, respectively. The wavevectors $\bk$ and $\bk'$ belong to a uniform Brillouin-zone grid with $N$ points.

By expanding the potential $V\big(\br;\{\tau_{\k\a p}\}\big)$ to first order in the ionic displacements from their equilibrium sites, and using the ladder operator notation for phonons, Eq.~\eqref{eq.soc.sq} can be written as:
  \begin{equation} \label{eq.soc.sq.displ}
  \hat{V}_{\rm SOC} = \hat{V}_{\rm SOC}^0 + \Vtot,
  \end{equation}
where $\hat{V}_{\rm SOC}^0$ is the operator evaluated with the ions in their equilibrium sites, and $\Vtot$ represents the phonon-assisted SOC Hamiltonian, which is responsible for the dynamic Rashba-Dresselhaus effect:
  \begin{equation} \label{eq.ph}
  \Vtot = N^{-\frac{1}{2}} \!\!\sum_{m n \nu, \bk, \bq} \sum_{s s'} g^{s' s}_{mn\nu}
   (\bk+\bq,\bk)\, \hcd_{m\bk+\bq s'}\hc_{n\bk s}(\ha_{\bq\nu}+\had_{-\bq\nu})~.
  \end{equation}
In this expression the ladder operators $\had_{\bq\nu}$/$\ha_{\bq\nu}$ create/annihilate a phonon with wavevector $\bq$, branch index $\nu$, and frequency $\w_{\bq\nu}$, and we introduced the spin-phonon matrix elements:
  \begin{equation} \label{eq.g}
  g_{mn\nu}^{s' s}  (\bk+\bq, \bk) = \prefac \< \psi_{m\bk+\bq s'}| \bs \cdot \nabla
  (\Delta_{\bq\nu} V) \times \bp\, | \psi_{n\bk s}\>~.
  \end{equation}
The quantity $\Delta_{\bq\nu} V$ in Eq.~\eqref{eq.g} represents the variation of the potential when the ions undergo a collective displacement along the phonon eigenmode $\bq\nu$; the notation employed here is the same as in Ref.~\onlinecite{Giustino2017}. The phonon wavevectors $\bq$ belong to the same uniform grid as the electron wavevectors $\bk$.

In Eq.~(2) of the main text we omitted electron band and phonon branch indices for notational simplicity, and we included spinor indices in the spin-phonon matrix elements as subscripts, $g_{s' s}(\bk+\bq, \bk)$.

\section{Spin-phonon matrix elements in the presence of time-reversal and inversion symmetry}\label{note-2}

Here we derive the transformation laws of the spin-phonon matrix elements resulting from time-reversal symmetry and inversion symmetry. In the presence of time-reversal symmetry, the vibrational eigenmodes can be chosen so as to fulfill the following relations, as shown in Ref.~\onlinecite{Maradudin1968}:
  \begin{equation} \label{eq.mode.trs}
  e_{\k\a,\nu}(-\bq) = e^*_{\k\a,\nu}(\bq)~.
  \end{equation}
As a consequence, the variation of the potential transforms as $\Delta_{\bq\nu} V^* = \Delta_{-\bq\nu} V$; using this relation inside Eq.~\eqref{eq.g} leads to:
  \begin{equation} \label{eq.matel.trs}
  g_{nm\nu}^{s's}(\bk',\bk) = g_{mn\nu}^{ss',*}(\bk,\bk')~.
  \end{equation}
In the presence of inversion symmetry, the vibrational eigenmodes satisfy the following property, as shown in Ref.~\onlinecite{Maradudin1968}: 
  \begin{equation}\label{eq.412}
  e_{\tilde \kappa \a,\nu}(-\bq) = -e^{i\varphi_{\bq\nu}}e_{\kappa \a,\nu}(\bq),
  \end{equation}
where the ion $\tilde \k$ is the inversion partner of $\k$, and $\varphi_{\bq\nu}$ is a phase to be determined by explicit calculations. By applying this transformation twice we obtain a relation for the phases:
  \begin{equation}\label{eq.phi}
   e^{i\varphi_{-\bq\nu}}  = e^{-i\varphi_{\bq\nu}}~.
  \end{equation}
As a result, the variation of the potential transforms as $[\nabla(\Delta_{\bq\nu} V)]^*(-\br) = e^{i\varphi_{\bq\nu}} \nabla(\Delta_{\bq\nu} V)(\br)$.

In the presence of both time-reversal symmetry and inversion symmetry, the unperturbed spinor states are fourfold degenerate, $\ve_{n\bk \bar{s}}=\ve_{n\bk s} = \ve_{n,-\bk \bar{s}}=\ve_{n,-\bk s}$, where $\bar{s}$ denotes the complement to $s$. The corresponding wavefunctions can be chosen in such a way as to satisfy the relations:
  \begin{equation}
  \psi_{n\bk \bar{s}}^{(1)}(\br) = -\psi_{n\bk s}^{(2),*}(-\br)~, \qquad
  \psi_{n\bk \bar{s}}^{(2)}(\br) = \psi_{n\bk s}^{(1),*}(-\br)~. \label{eq.kramers}
  \end{equation}
By using Eqs.~\eqref{eq.412} and \eqref{eq.kramers} inside Eq.~\eqref{eq.g} we obtain:
  \begin{equation}\label{eq.g.swap6}
  g_{mn\nu}^{\bar s' \bar s}(\bk+\bq, \bk) =  -(-1)^{s'-s}
  e^{-i\varphi_{\bq\nu}} g_{mn\nu}^{s' s,*}(\bk+\bq, \bk)~.
  \end{equation}
An additional relation can be derived by considering the transformation of spinors under inversion:
  \begin{equation}\label{eq.psi-par}
  \psi_{n,-\bk s}(-\br) = e^{i\chi_{n\bk s}}\,\psi_{n\bk s}(\br)~,
  \end{equation}
where $\chi_{n\bk s}$ is a phase that can be chosen arbitrarily when $\bk\ne 0$. In the main text we choose $\chi_{n\bk s}=0$. When $\bk=0$ this phase is set by the parity of the spinor, see Eq.~\eqref{eq.psi-phase} below. By combining Eqs.~\eqref{eq.g}, \eqref{eq.mode.trs}, \eqref{eq.412}, and \eqref{eq.psi-par} we find:
  \begin{equation}\label{eq.g.parity}
  g_{mn\nu}^{s' s}(-\bk-\bq, -\bk) =  -e^{i\varphi_{\bq\nu}} e^{-i\chi_{m\bk+\bq s'}} e^{i\chi_{n\bk s}}
  g_{mn\nu}^{s' s}(\bk+\bq,\bk)~.
  \end{equation}
We now apply these relations to the particular cases of electrons and phonons at the zone center.  In the case of zone-center phonons ($\bq=0$), time-reversal symmetry requires $e_{\k\a,\nu}(0)$ to be real-valued, as a result of Eq.~\eqref{eq.mode.trs}. In combination with inversion symmetry as expressed by Eq.~\eqref{eq.412}, this implies that also $e^{i\varphi_{0\nu}}$ is real-valued.  Therefore all zone-center phonons have defined parity:
  \begin{equation}\label{eq.phase}
   e_{\tilde \kappa \a,\nu}(0) = s_\nu \, e_{\kappa \a,\nu}(0)~,\qquad s_\nu = -e^{i\varphi_{0\nu}}=\pm 1~.
  \end{equation}
Using Eqs.~\eqref{eq.matel.trs}, \eqref{eq.g.swap6}, and \eqref{eq.phase} we obtain the following relations for the diagonal matrix elements of {\it even}-parity and {\it odd}-parity zone-center phonons:
  \begin{eqnarray}\label{eq.g.even1}
    \mbox{even} \hspace{10pt} s_\nu = +1: &&\qquad
        g_{nn\nu}^{22}(\bk,\bk)=g_{nn\nu}^{1 1}(\bk,\bk) \in \mathbb{R},\, \\
        &&\qquad
         g_{nn\nu}^{12}(\bk,\bk) = g_{nn\nu}^{21}(\bk,\bk) = 0~, \label{eq.g.even2} \\
    \mbox{odd} \hspace{13pt} s_\nu = -1: &&\qquad
      g_{nn\nu}^{22}(\bk,\bk)=-g_{nn\nu}^{1 1}(\bk,\bk) \in \mathbb{R},\,\label{eq.g.odd1} \\ &&\qquad
       g_{nn\nu}^{21}(\bk,\bk) = g_{nn\nu}^{12,*}(\bk,\bk)~.\label{eq.g.odd2}
  \end{eqnarray}
In the case of zone-center spinors ($\bk=0$), the matrix elements simplify further.  In fact, Eq.~\eqref{eq.psi-par} evaluated at $\bk=0$ gives $\psi_{n 0 s}(-\br) = e^{i\chi_{n 0 s}}\,\psi_{n 0 s}(\br)$. By using this relation twice (for $\br$ and for $-\br$) we see that $e^{i\chi_{n 0 s}}$ must be real-valued, therefore also all zone-center spinors have definite parity:
  \begin{equation}\label{eq.psi-phase}
  \psi_{n 0 s}(-\br) = p_{ns} \,\psi_{n 0 s}(\br)~,\qquad p_{n s} = e^{i\chi_{n 0 s}}=\pm 1~.
  \end{equation}
Furthermore, by combining this result with Eq.~\eqref{eq.kramers}, we find that the two Kramers-degenerate spinors must have the same parity:
  \begin{equation}\label{eq.psi-phase2}
    p_{n \bar s} = p_{ns}~.
  \end{equation}
By using Eqs.~\eqref{eq.phase}, \eqref{eq.psi-phase}, and \eqref{eq.psi-phase2} inside Eq.~\eqref{eq.g.parity}, we obtain:
  \begin{equation}\label{eq.g.odd.zero}
    g_{nn\nu}^{s' s}(0,0) = s_\nu \,g_{nn\nu}^{s' s}(0,0)~.
  \end{equation}
Therefore in the case of {\it odd}-parity phonons ($s_\nu=-1$) the spin-phonon matrix elements for $\bk=\bq=0$ must vanish altogether.

\section{Fock-space perturbation theory at zero temperature}\label{note-3}

We derive the first- and second-order corrections to the electron energies arising from the spin-phonon coupling of Eq.~\eqref{eq.ph}.

We start by considering the Fock state $|n\bk s,0\>$ obtained from the phonon vacuum and Fermi vacuum $|0\>$ by creating an extra electron in the conduction band, $|n\bk s,0\> = \hcd_{n\bk s}|0\>$.  The Fermi vacuum corresponds to filled valence bands and empty conduction bands of an insulator or semiconductor. The Hamiltonian in the absence of spin-phonon coupling is:
  \begin{equation}\label{eq.H0}
  \hH_0 = \sum_{n\bk s} \ve_{n\bk s} \,\hcd_{n\bk s} \hc_{n\bk s}
  + \sum_{\bq\nu} \hbar\w_{\bq\nu}\,\had_{\bq\nu}\ha_{\bq\nu}~.
  \end{equation}
This expression implicitly assumes the quasiparticle approximation for electrons and the harmonic approximation for lattice dynamics. The energies $\ve_{n\bk s}$ already include electron-electron interactions and the static spin-orbit coupling contribution $\hat{V}_{\rm SOC}^0$ from Eq.~\eqref{eq.soc.sq.displ}.  The zero-point energy adds an inessential constant to the total energy, and is omitted in Eq.~\eqref{eq.H0}.

We emphasize that the eigenstates of $\hH_0$ are stationary states only when electron-phonon interactions are absent, i.e. in the unperturbed system. For these states the absence of spin-splitting can be derived using elementary symmetry arguments. Conversely, in the interacting electron-phonon system, quasiparticle states are not eigenstates of the Hamiltonian, therefore elementary symmetry arguments do not apply. In the interacting case the possibility of spin splitting must be analyzed in terms of many-body perturbation theory.

In the presence of time-reversal and inversion symmetry, Kramers' degeneracy implies $\ve_{n\bk \bar{s}}=\ve_{n\bk s}=\ve_{n,-\bk \bar{s}}=\ve_{n,-\bk \bar{s}}$, therefore to obtain the first-order corrections to the energies we employ degenerate perturbation theory. The corrections are obtained as the eigenvalues of the perturbation matrix evaluated in the degenerate subspace.  Using Eqs.~\eqref{eq.ph} we find:
  \begin{eqnarray}\label{eq.corr.one.a}
  \< n\bk' s',0 | \Vtot | n\bk s,0\> = 0, \qquad \mbox{for any  } s,s',\bk'=\pm \bk~,
  \end{eqnarray}
since the ladder operators change the number of phonons but the bra and ket correspond both to the phonon vacuum. Therefore the first-order correction to the eigenvalues vanishes identically. The same reasoning applies to all odd ($2n+1$) orders  in perturbation theory, since at least one among the $2n+1$ matrix elements in the product must vanish.

Since the spin degeneracy is not lifted at the first order, we proceed to the second order.  If we have states $|u\>$ and $|v\>$ which are degenerate to zeroth and first order, and which are not degenerate with any other state $|i\>$, then the matrix to be diagonalized for obtaining the eigenvalue shifts is:\cite{Landau1981} 
  \begin{equation}\label{eq.secord}
  \sum_{i\ne u,v}\frac{\< u | \Delta \hat{H}|i\>\<i| \Delta \hat{H}| v \>}{\ve_u-\ve_i}~,
  \end{equation}
where $\Delta \hat{H}$ stands for the perturbation Hamiltonian. In the present case this expression takes the form:
  \begin{equation} \label{eq.delta2.nondeg}
  V_{n\bk's',n\bk s} = {\sum_i}^{\,\prime} \,\frac{ \< n\bk s',0 | \Vtot| i \> \<i|\Vtot|n\bk s,0\>
  }{\ve_{n\bk s}-\ve_i}~,
  \end{equation}
where the prime in the summation means that $|i\> \ne |n\bk s,0\>, |n\bk \bar{s},0\>, |n,-\bk s,0\>, |n,-\bk \bar{s},0\>$. From inspection of Eq.~\eqref{eq.ph} we observe that the only nonzero matrix elements must be of the form:
  \begin{equation} \label{eq.matel.a}
  \<m\bk+\bq s'',1_{-\bq\nu}|\Vtot|n\bk s,0\> = N^{-\frac{1}{2}}\, g_{mn\nu}^{s'' s}(\bk+\bq,\bk)~.
  \end{equation}
The energy of the corresponding virtual state is $\ve_i = \ve_{m\bk+\bq s''} + \hbar\w_{\bq\nu}$.
In reality this choice is a simplification, because we are neglecting the possibility of creating holes in the valence bands, which leads to additional terms related to Pauli blocking and vacuum polarization. These additional terms are discussed in Supplementary Note~\ref{note-5}. After combining Eqs.~\eqref{eq.delta2.nondeg}, \eqref{eq.matel.a}, and \eqref{eq.matel.trs} we obtain:
  \begin{equation}\label{eq.all.v}
  V_{n\bk's',n\bk s} = \d_{\bk,\bk'}\frac{1}{N}\sum_{m,\bq\nu} \frac{\sum_{s'\!'}
  g_{nm\nu}^{s' s'\!'}(\bk,\bk+\bq) g_{mn\nu}^{s'\!' s}(\bk+\bq,\bk)}
  {\ve_{n\bk} -\ve_{m\bk+\bq}-\hbar\w_{\bq\nu}}~,
  \end{equation}
where we have omitted the spinor label in the eigenenergies as $\ve_{n\bk s}=\ve_{n\bk \bar{s}}$.  Now we can use Eqs.~\eqref{eq.phi} and \eqref{eq.g.swap6} in Eq.~\eqref{eq.all.v}, and the fact that $\Vtot$ is an Hermitian operator, to find:
  \begin{equation}
  V_{\bar{s} \bar s} = V_{ss}~, \qquad V_{\bar{s} s} = 0~.
  \end{equation}
As a result, the second-order corrections to the eigenvalues are:
  \begin{equation} \label{eq.nosplitting}
  \Delta \ve_{n\bk\,1 } = \Delta \ve_{n\bk\, 2} = \Delta \ve_{n\bk}~,
  \end{equation}
having defined:
  \begin{equation} \label{eq.shift}
  \Delta \ve_{n\bk} = \frac{1}{N}\sum_{m,\bq\nu} \frac{ |g_{mn\nu}^{1 1}(\bk,\bq)|^2
  +|g_{mn\nu}^{1 2}(\bk,\bq)|^2 } {\ve_{n\bk} -\ve_{m\bk+\bq}-\hbar\w_{\bq\nu}}~.
  \end{equation}
This result indicates that the phonon-assisted spin-orbit coupling $\Vtot$ does not lift the Kramers' degeneracy to second order in perturbation theory at zero temperature. We also note that, unlike the standard energy correction in spin-unpolarized calculations,\cite{Giustino2017} Eq.~\eqref{eq.shift} involves both same-spin and opposite-spin matrix elements.

\section{Evaluation of spin splitting to fourth order in spin-phonon coupling}\label{note-4}

As an example we sketch the evaluation of higher-order perturbative corrections to the energy at zero temperature. At fourth order, Eq.~\eqref{eq.secord} generalizes to:
  \begin{equation}
  V^{(4)}_{s's} = {\sum_{ijk}}^{\,\prime} \frac{\< n\bk s',0 | \Vtot|k\> \<k| \Vtot| j \>
    \<j| \Vtot| i \>  \<i| \Vtot| n\bk s, 0\>}{(\ve_{n\bk}-\ve_k)
       (\ve_{n\bk}-\ve_j) (\ve_{n\bk}-\ve_i) }~,
  \end{equation}
where the prime indicates that the states $| n\bk s, 0\>$ and $| n,-\bk s, 0\>$ with $s,s'=1,2$ are excluded from the summation.  It is convenient to enumerate the various possibilities for the intermediate states using diagrams. In Fig.~S\ref{fig-s2} we show the inequivalent fourth-order virtual scattering diagrams, and we indicate the spin and momenta for the case A. The resulting expression for this diagram is:
  \begin{eqnarray}\label{eq.4th}
  &&V^{(4,A)}_{s's} = N^{-2}\!\!\!\!\!\!\!
   \sum_{\{n_i\},\{\nu_i\},\{s_i\}}
     g_{ns' ,n_3s_3,\nu_2}(\bk,\bk+\bq')
     g_{n_3s_3, n_2s_2,\nu_1}(\bk+\bq',\bk+\bq+\bq') \nonumber \\
   &&
   \times\, g_{n_2s_2, n_1s_1,\nu_2}(\bk+\bq+\bq',\bk+\bq)
       g_{n_1 s_1,n s,\nu_1}(\bk+\bq,\bk) \nonumber \\
  &&
    \times\,\,
   \frac{ 1 }{
       [\ve_{n\bk}- (\ve_{n_3\bk+\bq}+\hbar\w_{-\bq\nu_2})]
       [\ve_{n\bk}- (\ve_{n_2\bk+2\bq}+\hbar\w_{-\bq\nu_1}+\hbar\w_{-\bq\nu_2})]
       [\ve_{n\bk}- (\ve_{n_1\bk+\bq}+\hbar\w_{-\bq\nu_1})] }~. \nonumber\\
  \end{eqnarray}
By applying the transformation law in Eq.~\eqref{eq.g.swap6} to this expression we find:
    \begin{eqnarray}
  &&V^{(4,A)}_{\bar s' \bar s} = N^{-2}\!\!\!\!\!\!\!
   \sum_{\{n_i\},\{\nu_i\},\{s_i\}} (-1)^{s'-s_3+s_3-s_2+s_2-s_1+s_1-s}
        e^{-i\varphi_{-\bq\nu_2}}
     e^{-i\varphi_{-\bq\nu_1}} e^{-i\varphi_{\bq\nu_2}} e^{-i\varphi_{\bq\nu_1}} \nonumber \\
     &&
     \times\,\,g^*_{n  s' ,n_3s_3\,nu_2}(\bk,\bk+\bq')
     g^*_{n_3s_3, n_2s_2,\nu_1}(\bk+\bq',\bk+\bq+\bq') \nonumber \\
   &&\times\,
    g^*_{n_2s_2, n_1s_1,\nu_2}(\bk+\bq+\bq',\bk+\bq)    g^*_{n_1 s_1,n s,\nu_1}(\bk+\bq,\bk) \nonumber \\
  &&
    \times\,\,
   \frac{ 1 }{
       [\ve_{n\bk}- (\ve_{n_3\bk+\bq}+\hbar\w_{-\bq\nu_2})]
       [\ve_{n\bk}- (\ve_{n_2\bk+2\bq}+\hbar\w_{-\bq\nu_1}+\hbar\w_{-\bq\nu_2})]
       [\ve_{n\bk}- (\ve_{n_1\bk+\bq}+\hbar\w_{-\bq\nu_1})] }~. \nonumber \\
  \end{eqnarray}
After simplifying the exponents we have:
  \begin{equation}
    V^{(4,A)}_{\bar s' \bar s} =  (-1)^{s'-s} V^{(4,A)\,*}_{s's}~.
  \end{equation}
This relation implies $V^{(4,A)\,*}_{21} = -V^{(4,A)}_{12}$.  Since the operator $\Vtot$ is Hermitian, we also have $V^{(4,A)\,*}_{21} = V^{(4,A)}_{12}$. Therefore $V^{(4,A)}_{12}=V^{(4,A)}_{21}=0$. The same reasoning holds for diagrams B, C, and D of Fig.~S\ref{fig-s2}, and for all higher even orders in perturbation theory.

\section{Fock-space perturbation theory at finite temperature}\label{note-5}

We determine the energy renormalization from spin-phonon interaction to second order at finite temperature.  We consider the state $| n\bk s,P\> = \hcd_{n\bk s}|P\>$, where $|P\>$ represents the Fermi vacuum with phonons excited to the state of energy $E_P$. For example the state $|P\> = \had_{\bq\,1}\hat{a}^{\dagger 2}_{\bq\,2}|0\>$ is an eigenstate of the unperturbed system with eigenenergy $E_P = \hbar\w_{\bq\,1}+2\hbar\w_{\bq\,2}$. It can be verified that Eq.~\eqref{eq.corr.one.a} remains valid also for the state $| n\bk s,P\>$:
  \begin{eqnarray}\label{eq.corr.one.a.2}
  \< n\bk s',P | \Vtot | n\bk s,P\> = 0, \qquad \mbox{for any  } s,s'~,
  \end{eqnarray}
because $\Vtot$ only couples states which differ by one phonon in each mode.

The second-order correction is evaluated in the same way as for the zero-temperature case. Eq.~\eqref{eq.delta2.nondeg} remains almost unchanged, except that we replace $|n\bk s,0\>$ by $|n \bk s,P\>$ and we skip the states $|n\bk s,P\>, |n\bk \bar{s},P\>$ in the summation.

The matrix elements $\<i|\Vtot|n\bk s,P\>$ in Eq.~\eqref{eq.delta2.nondeg} vanish unless the phonon component of $|i\>$ contains $|P\pm 1_{\bq\nu}\>$. If we denote by $|i_e\>$ the electronic component of the state $|i\>$ and $n_{\bq\nu}$ the number of phonons in mode $\bq\nu$ contained in $|P\>$, using the properties of the ladder operators we can write:
  \begin{equation}\label{eq.matel.T.1}
  \<i|\Vtot|n\bk s,P\> = N^{-\frac{1}{2}}
  \<i_e| \hcd_{m'\bk'+\bq s'''}\hc_{n'\bk' s''} |n\bk s\>
  \sqrt{n_{\bq\nu}+\frac{1}{2}\pm \frac{1}{2}}\,\,
   g^{s''' s''}_{m'n'\nu}(\bk'+\bq,\bk')~.
  \end{equation}
Unlike the simplified derivations in Supplementary Notes~\ref{note-3} and \ref{note-4}, now we consider the most general case where holes can be excited out of the valence bands. Let $a = (n\bk s)$, $b = (n'\bk' s'')$, and $c=(m'\bk'+\bq s''')$; we must distinguish four cases depending on the relations between $a$, $b$, and $c$.  (A) If $b=c$ then the matrix element is nonzero only when $\bq=0$, but the weight of this contribution vanishes in the limit of dense Brillouin zone sampling since $1/N\rightarrow 0$. If $b\ne c$ there are three possibilities: (B) $a=c$, (C) $a=b$, and (D) $a$ differs from $b$ and $c$. In case (B) the Dirac braket vanishes identically.  In case (C) the braket yields $(1-f_{m'\bk'+\bq s'''})f_{n\bk s}$ and the virtual state is $|i_e\> = |1_{m'\bk'+\bq s'''}\>$ with energy $E_i = \ve_{m'\bk'+\bq}\pm \hbar\w_{\bq\nu}$. In case (D) the braket becomes $(1-f_{m'\bk'+\bq s'''})f_{n'\bk' s''}$ and the virtual state is $|i_e\> = |1_{n\bk s}-1_{n'\bk' s''}+1_{m'\bk'+\bq s'''}\>$ with energy $E_i = \ve_{n\bk}-\ve_{n'\bk'}+\ve_{m'\bk'+\bq} \pm \hbar\w_{\bq\nu}$.  The processes (A)-(D) are illustrated schematically in Fig.~S\ref{fig-s3}.  Only processes of type (C) and (D) contribute to Eq.~\eqref{eq.delta2.nondeg}, therefore we proceed to replace the matrix elements and energies for these processes inside Eq.~\eqref{eq.delta2.nondeg}. We find:
  \begin{equation}\label{eq.type3}
  V^{\rm (C)}_{s's}(n\bk) = f_{n\bk s}\frac{1}{N}\sum_{m s'',\bq\nu}
  g_{mn\nu}^{s'' s',*}(\bk+\bq,\bk) g_{mn\nu}^{s'' s}(\bk+\bq,\bk) \sum_{r=\pm 1}
  \frac{[(1+r)/2+n_{\bq\nu}](1-f_{m\bk+\bq s''})} {\ve_{n\bk}-\ve_{m\bk+\bq}-r\hbar\w_{\bq\nu}}~.
  \end{equation}
  \begin{equation}\label{eq.type4}
  V^{\rm (D)}_{s's}(n\bk) = \d_{ss'}\sum_{n''\bk''s''} (1-\d_{n\bk s,n''\bk''s''})
  V^{\rm (C)}_{s's}(n''\bk'')~.
 \end{equation}
This last contribution corresponds to the generation of virtual electron-hole pairs across the gap via phonon absorption or the annihilation of pairs via phonon emission (Fig.~S\ref{fig-s3}). These processes are the vacuum fluctuations of quantum field theory.\cite{Schrieffer1983}

It is convenient to rewrite $V^{\rm (D)}_{s's}$ as follows.  Equation~\eqref{eq.type4} does not include virtual transitions into the conduction state $n\bk s$, as a consequence of Pauli blocking. This is seen in Eq.~\eqref{eq.type3} by noting that the term $(1-f_{m\bk+\bq s''})$ vanishes when $m\bk+\bq s''=n\bk s$. In order to single out this effect, we add and subtract this virtual transition from Eq.~\eqref{eq.type4}. After re-labelling the variables, and using Eq.~\eqref{eq.matel.trs} to recast the matrix elements in a more convenient form, we obtain:
   \begin{equation}\label{eq.pauli}
   V^{\rm (D)}_{ss}(n\bk) = \Delta E_{\rm gs} + \frac{1}{N} \sum_{m s'',\bq\nu}
   |g_{mn\nu}^{s'' s}(\bk+\bq,\bk)|^2 \sum_{r=\pm 1} \frac{[(1-r)/2+n_{\bq\nu}]f_{m\bk+\bq s''}}
   {\ve_{n\bk}-\ve_{m\bk+\bq}-r\hbar\w_{\bq\nu}}~, \quad V^{\rm (D)}_{\bar{s}s}(n\bk)=0~.
   \end{equation}
Here $\Delta E_{\rm gs}$ represents the correction to the ground state energy (i.e.\ the valence electrons) resulting from spin-phonon interactions; this term corresponds to Eq.~\eqref{eq.type4} with the Pauli-blocked transition enabled.  The remainder is minus the contribution of the blocked transition. In quantum field theory the energy shift $\Delta E_{\rm gs}$ is referred to as the vacuum polarization.\cite{Mahan1993,Scalapino1969,Grimvall1981,Schrieffer1983} This shift is the same for every electronic state, therefore it does not affect the following discussion.  After adding Eqs.~\eqref{eq.type3} and \eqref{eq.pauli} we obtain the perturbation matrix:
  \begin{eqnarray}\label{eq.pauli.all.1}
  V_{\bar{s}s}(n\bk) &=& \frac{1}{N}\sum_{m s'',\bq\nu}
  g_{mn\nu}^{s'' \bar{s},*}(\bk+\bq,\bk)g_{mn\nu}^{s'' s}(\bk+\bq,\bk) \sum_{r=\pm 1}
  \frac{[(1+r)/2+n_{\bq\nu}](1-f_{m\bk+\bq s''})} {\ve_{n\bk}-\ve_{m\bk+\bq}-r\hbar\w_{\bq\nu}},\qquad\\
  V_{s s}(n\bk) &=& \frac{1}{N}\sum_{m s'',\bq\nu} |g_{mn\nu}^{s'' s}(\bk+\bq,\bk)|^2
  \sum_{r=\pm 1} \frac{1/2+ r(1-2f_{m\bk+\bq s''})/2+ n_{\bq\nu}}
  {\ve_{n\bk}-\ve_{m\bk+\bq}-r\hbar\w_{\bq\nu}} + \Delta E_{\rm gs}~.\qquad \label{eq.pauli.all.2}
  \end{eqnarray}
In order to take the system temperature into account, we perform a canonical average of this matrix over all possible phonon occupation numbers.  Since Eqs.~\eqref{eq.pauli.all.1}-\eqref{eq.pauli.all.2} are linear in $n_{\bq\nu}$, the problem is equivalent to carrying out the canonical average of the mean-square vibrational amplitudes of the quantum harmonic oscillator. The result of this textbook procedure is that the integers $n_{\bq\nu}$ appearing in Eqs.~\eqref{eq.pauli.all.1}-\eqref{eq.pauli.all.2} must be replaced by the corresponding Bose-Einstein occupations at the temperature $T$, $n_{\bq\nu}(T)$.

Now we check whether the perturbation admits off-diagonal components. Using Eq.~\eqref{eq.pauli.all.1} and Eq.~\eqref{eq.g.swap6} we find:
  \begin{equation}\label{eq.offdiag-check}
  V_{21}(n\bk) = \frac{1}{N} \sum_{m,\bq} (f_{m\bk+\bq\,2}-f_{m\bk+\bq\,1})
  \sum_{\nu,r=\pm 1} g_{mn\nu}^{12,*}(\bk+\bq,\bk) g_{mn\nu}^{11}(\bk+\bq,\bk)
   \frac{[(1+r)/2+n_{\bq\nu}]}
  {\ve_{n\bk}-\ve_{m\bk+\bq}-r\hbar\w_{\bq\nu}}~.
  \end{equation}
The contributions to this sum arising from valence states cancel out since $f_{m\bk+\bq\,1}= f_{m\bk+\bq\,2}=1$. For all conduction states except $n\bk s$ there is also cancellation since $f_{m\bk+\bq\,1}=f_{m\bk+\bq\,2}=0$. Therefore only the transition with $m=n$ and $\bq=0$ contributes to the sum over $m,\bq$, and Eq.~\eqref{eq.offdiag-check} becomes $V_{21}(n\bk) =N^{-1}(-1)^s \sum_{\nu} g_{nn\nu}^{12,*}(\bk,\bk) g_{nn\nu}^{11}(\bk,\bk)/\hbar\w_{0\nu}$. In the limit of dense Brillouin-zone sampling we have $V_{21}=0$ since $1/N\rightarrow 0$.

The correction to the energy resulting from the diagonal elements of $V_{ss'}$ is given by Eq.~\eqref{eq.pauli.all.2}:
  \begin{eqnarray}
  \Delta \ve_{n\bk s'}(T) &=&
  \frac{1}{N}{\sum_{m\bq\nu}}^{\prime}
  \sum_{s''}|g_{mn\nu}^{s'' s'}(\bk+\bq,\bk)|^2
  \left[\frac{1+n_{\bq\nu}(T)-f_{m\bk+\bq}}
  {\ve_{n\bk}-\ve_{m\bk+\bq}-\hbar\w_{\bq\nu}} +\frac{n_{\bq\nu}(T)+f_{m\bk+\bq}}
  {\ve_{n\bk}-\ve_{m\bk+\bq}+\hbar\w_{\bq\nu}}\right]
  \nonumber \\
   &+& \Delta E_{\rm gs}
   -\frac{1}{N}\sum_{\nu} \sum_{s''}\frac{|g_{nn\nu}^{s'' s'}(\bk,\bk)|^2 }
   {\hbar\w_{0\nu}}
   + \frac{2}{N}\sum_{\nu} \frac{|g_{nn\nu}^{s s'}(\bk,\bk)|^2}{\hbar\w_{0\nu}}.~\label{eq.shift.T}
  \end{eqnarray}
Here the primed summation runs over all transitions except those where the virtual electronic state is $|n\bk\,1\>$ or $|n\bk\,2\>$, and the extra electron has been added to the spinor state $n\bk s$. The last two terms are for spin-flip transitions within the same shell, and vanish in the limit of dense Brillouin-zone sampling since they only involve $\bq=0$ phonons and $1/N\rightarrow 0$. The first line of Eq.~\eqref{eq.shift.T} is finite in the limit of dense sampling, but is independent of $s'$, therefore the energies of the Kramers' degenerate spinors undergo the same shift. The evaluation of higher orders in perturbation theory proceeds along the same lines, and leads to the same result.

\section{Spin-splitting for out-of-equilibrium coherent states}\label{note-6}

In this section we analyze the effect of spin-phonon coupling on a coherent state.\cite{Tannoudji1977} Coherent states are non-stationary since they are not eigenstates of the unperturbed Hamiltonian in Eq.~\eqref{eq.H0}, therefore in the following we focus on a snapshot of the many-body wavefunction at a given time $t$.

Coherent states for one extra electron in the conduction band and for the phonon $\bq\nu$, normalized within the BvK supercell, can be constructed as follows:\cite{Tannoudji1977}
  \begin{equation}\label{eq.coh}
  |n\bk s;\bq\nu,u\> = \begin{cases} \bq\ne 0: & \hcd_{n\bk s}
  \exp(-u^2 N/2) \exp\Big[u\, (N/2)^{1/2}\, ( \had_{\bq\nu}+\had_{-\bq\nu})\Big]|0\>, \\[10pt]
  \bq= 0: & \hcd_{n\bk s} \exp(-u^2 N/2) \exp\Big[u\, N^{1/2}\, \displaystyle\had_{\bq\nu}\Big]|0\>~,
  \end{cases}
  \end{equation}
where $u$ is a real number and the combination of $\pm \bq$ when $\bq\ne 0$ is needed to generate a standing wave. The prefactor $N^{1/2}$ takes care of the normalization in the BvK supercell. $|n\bk s;\bq\nu,u\>$ is an eigenstate of the phonon annihilation operators:
  \begin{equation} \label{eq.coh.2}
  \ha_{\bq\nu} |n\bk s;\bq\nu,u\> = \ha_{-\bq\nu} |n\bk s;\bq\nu,u\> = \begin{cases}
  \bq\ne 0: & u (N/2)^{1/2} |n\bk s;\bq\nu,u\> \\[3pt] \bq = 0: & u N^{1/2} |n\bk s;\bq\nu,u\>~.
  \end{cases}
  \end{equation}
Using Eq.~\eqref{eq.coh.2}, the relation between ionic displacements and ladder operators,\cite{Giustino2017} and the time-reversal symmetry of vibrational eigenmodes from Eq.~\eqref{eq.mode.trs}, the expectation value of the ionic displacements in the state $|n\bk s;\bq\nu,u\>$ can be written as:
  \begin{equation}\label{eq.displ.coh}
  \<n\bk s;\bq\nu,u| \Delta\tau_{\kappa\a p}|n\bk s;\bq\nu,u\> =
  \begin{cases} \bq \ne 0: & u \,(\hbar/2 M_\k \w_{\bq\nu})^{1/2}
  \,2\sqrt{2} \,{\rm Re}\!\left[ e^{i\bq\cdot\bRp} e_{\kappa\a,\nu}(\bq)\right]~, \\
  \bq = 0: & u\, (\hbar/2 M_\k \w_{\bq\nu})^{1/2}\, 2\,e_{\kappa\a,\nu}(\bq)~.  \end{cases}
  \end{equation}
where $M_\k$ is the mass of ion $\k$ and $\bRp$ is the direct lattice vector pointing to the unit cell $p$ in the BvK supercell. Equation~\eqref{eq.displ.coh} shows that the $u$ sets the magnitude of the ionic displacements from their equilibrium sites.  Using Eq.~\eqref{eq.coh.2}, the expectation value of the unperturbed Hamiltonian of Eq.~\eqref{eq.H0} for this coherent state is found to be:
  \begin{equation}
  \<n\bk s;\bq\nu,u| \hH_0 |n\bk ;\bq\nu,u\> = \ve_{n\bk s} + u^2 N\,\hbar\w_{\bq\nu}~.
  \end{equation}
This is the total energy in the BvK supercell, except the zero point energy which has been omitted in Eq.~\eqref{eq.coh.2}. We see that the state $|n\bk ;\bq\nu,u\>$ contains the equivalent of $u^2$ phonons of energy $\hbar\w_{\bq\nu}$ per unit cell of the crystal.

Since the coherent state is not an eigenstate of $\hH_0$, we cannot employ perturbation theory as in Supplementary Notes~\ref{note-3} and \ref{note-5}. Nevertheless we can evaluate the expectation value of the total Hamiltonian $\hat{H}_{\rm tot} =\hH_0+\Vtot$ on $|n\bk ;\bq\nu,u\>$.  After defining $H_{{\rm tot},s s'}= \<n\bk s';\bq\nu,u| \hat{H}_{\rm tot} |n\bk s;\bq\nu,u\>$ and combining Eqs.~\eqref{eq.ph}, \eqref{eq.H0}, \eqref{eq.coh}, and \eqref{eq.coh.2}, we obtain:
  \begin{equation}\label{eq.expt.values.2}
  H_{{\rm tot},s s'} = \d_{s's}\left[\ve_{n\bk} + u^2 N \hbar\w_{\bq\nu}
   + \d_{\bq,0}\,2 u \!\sum_{n' \bk'\ne n \bk}\sum_{s''}\, g^{s'' s''}_{n'n'\nu}
   (\bk',\bk') \right]
  + \d_{\bq,0}(1-\d_{ss'})\,2 u \, g^{s' s}_{nn\nu}(\bk,\bk)~.
  \end{equation}
This expression corresponds to the energy of a BvK supercell; accordingly the electronic contribution does not scale with $N$ since we are considering one electron per supercell, while the vibrational contribution scales with $N$ because there are $u^2$ phonons in each unit cell. The Kronecker delta $\d_{\bq,0}$ indicates that only $\bq=0$ phonons can generate a non-vanishing energy shift. This is consistent with the fact that only these phonons can lift the inversion symmetry of the entire crystal. Accordingly, in the following we consider only $\bq=0$ states.

We note that the condition $\bq=0$ is derived here for infinitely extended bulk crystals.  This condition can be relaxed by considering crystals of finite size.  In fact the Rashba-Dresselhaus spin-splitting discussed below is allowed whenever the phonon wavelength exceeds the linear size $L$ of the sample, so that the average of the atomic displacements over the crystal is non-vanishing. Therefore the criterion to be fulfilled by the coherent state is $|\bq|<\pi/L$.

Equation~\eqref{eq.expt.values.2} can be simplified by using time-reversal and inversion symmetry. We distinguish the cases of odd-parity and even-parity phonons:

\vspace{5pt}
\textit{Odd-parity zone-center phonon}

Using Eqs.~\eqref{eq.g.odd1}-\eqref{eq.g.odd2} inside Eq.~\eqref{eq.expt.values.2} we obtain:
  \begin{equation}\label{eq.expt.values.4}
  H_{{\rm tot},s s'} = \begin{bmatrix} E_0+ \Delta_1 & \Delta_2 \\ \Delta_2^* & E_0 -\Delta_1
  \end{bmatrix}~,
  \end{equation}
having defined:
  \begin{equation}\label{eq.delta}
  E_0 = \ve_{n\bk} + u^2 N_p \hbar\w_{\bq\nu}, \qquad \Delta_1 = \d_{\bq,0}\,2 u\, g^{1 1}_{nn\nu}(\bk,\bk),
   \qquad
  \Delta_2 = \d_{\bq,0}\,2 u\, g^{1 2}_{nn\nu}(\bk,\bk)~.
  \end{equation}
The eigenvalues of the matrix are:
  \begin{equation}
  E = E_0 \pm \sqrt{\Delta_1^2 + |\Delta_2|^2}~.
  \end{equation}
If we take the coherent states defined in Eq.~\eqref{eq.coh} as approximations to the true many-body states of the unperturbed system, then these eigenvalues represent the excitation energies in the presence of spin-phonon coupling. In this case the Kramers' degeneracy is lifted and the energy splitting is:
  \begin{equation}\label{eq.split.coh}
  |\ve_{n\bk\,1}-\ve_{n\bk\,2}| = 4 |u| \sqrt{\vphantom{I^-}
  |g^{1 1}_{nn\nu}(\bk,\bk)|^2 + |g^{1 2}_{nn\nu}(\bk,\bk)|^2}~.
  \end{equation}

\textit{Even-parity zone-center phonon}

Using Eqs.~\eqref{eq.g.even1}-\eqref{eq.g.even2} inside Eq.~\eqref{eq.expt.values.2} we obtain:
 \begin{equation}\label{eq.expt.values.5}
  H_{{\rm tot},s s'} = \begin{bmatrix} E_0+ \Delta_3 & 0 \\ 0 & E_0 +\Delta_3
  \end{bmatrix}~,
  \end{equation}
having defined:
  \begin{equation}\label{eq.delta3}
  \Delta_3 = \d_{\bq,0}\,\,4 u \!\!\sum_{n\bk'\ne n\bk} g^{1 1}_{n'n'\nu}(\bk',\bk')~.
  \end{equation}
Since this contribution is diagonal in the spinor labels, coherent even-parity phonons do not lead to spin-splitting.

\section{Calculation of Rashba energy for coherent states} \label{note-7}

We outline a procedure to determine the Rashba energy from first principles calculations.  We consider a density functional theory (DFT) calculation, where the ionic degrees of freedom are described within the classical and adiabatic approximations. If we displace the ions in every unit cell according to the expectation values of the coherent state in
Eq.~\eqref{eq.displ.coh}:
  \begin{equation}\label{eq.displ.dft}
  \Delta\tau^{\rm DFT}_{\kappa\a}= 2 u \,(\hbar/2 M_\k \w_{0\nu})^{1/2} e_{\kappa\a,\nu}(0)~,
  \end{equation}
then it is straightforward to show that, for an odd-parity phonon, the Kramers' degenerate states split as follows (to first order in perturbation theory):
  \begin{equation}\label{eq.split.dft}
  \Delta \ve^{\rm DFT}_{n\bk} = \pm 2 |u| \sqrt{\vphantom{I^-}
  |g^{1 1}_{nn\nu}(\bk,\bk)|^2 + |g^{1 2}_{nn\nu}(\bk,\bk)|^2}~.
  \end{equation}
This result is identical to the spin splitting found for coherent states in Eq.~\eqref{eq.split.coh}. Therefore, by performing standard DFT calculations with finite atomic displacements, we can mimic the effect of a coherent state.

\noindent
The long-wavelength expansion of the matrix elements appearing in Eq.~\eqref{eq.split.dft} is:
\begin{equation}\label{eq.g.expansion}
  g_{nn\nu}^{s' s}(\bk, \bk) = g_{nn\nu}^{s's}(0,0) + \bk \cdot \nabla_\bk
 \left.g_{nn\nu}^{s' s}(\bk, \bk)\right|_{\bk = 0}~.
\end{equation}
Since we are considering a system which is invariant under time-reversal and inversion symmetry in the presence of a coherent odd-parity phonon, from Eq.~\eqref{eq.g.odd.zero} we have $g_{nn\nu,0}^{s's}=0$. Therefore by replacing Eq.~\eqref{eq.g.expansion} inside Eq.~\eqref{eq.split.dft} we find:
  \begin{equation}\label{eq.de}
  \Delta \ve^{\rm DFT}_{n\bk} = \pm 2 \,|u|\,|C_{n\nu}(\bm{\hat{k}})|\,\hspace{0.3pt}k~,
  \end{equation}
where we have defined the dynamic Rashba-Dresselhaus coupling constant $C_{n\nu}(\bm{\hat{k}})$ as: \begin{equation}\label{eq.C.rashba}
  C_{n\nu}^2(\bm{\hat{k}}) =
  | \bm{\hat{k}} \cdot [\nabla_\bk \, g_{nn\nu}^{11}(\bk, \bk)]_{\bk=0}
  |^2 + | \bm{\hat{k}} \cdot [\nabla_\bk \, g_{nn\nu}^{12}(\bk,\bk)]_{\bk=0} |^2~,
  \end{equation}
and $\bm{\hat{k}}$ is the unit vector in the direction of $\bk$.  In the case of a parabolic conduction band minimum with effective mass $m^*$, the correction $\Delta \ve^{\rm DFT}_{n\bk}$ of Eq.~\eqref{eq.de} leads to the standard Rashba-Dresselhaus band splitting of the renormalized energies
$E_{n\bk} = \ve_{n\bk}+\Delta \ve^{\rm DFT}_{n\bk}$:
  \begin{equation}\label{eq.dyn.rashba1}
    E_{n\bk} = \frac{\hbar^2 (k\pm k_{\rm R})^2}{2m^*} - E_{\rm R}~,
  \end{equation}
with the Rashba wavevector and energy for the electron band $n$, phonon branch $\nu$, and direction $\bm{\hat{k}}$ are given by:
  \begin{equation}\label{eq.dyn.rashba2}
    k_{\rm R} = |u| \frac{2m^* |C_{n\nu}(\bm{\hat{k}})|}{\hbar^2}~,\qquad
    E_{\rm R} = \frac{2m^*}{\hbar^2 } u^2 C^2_{n\nu}(\bm{\hat{k}})~.
  \end{equation}

\section{Coupling to long-wavelength longitudinal-optical phonons in polar crystals} \label{note-8}

Here, we evaluate the dynamic Rashba-Dresselhaus spin-splitting of Eqs.~\eqref{eq.dyn.rashba1}-\eqref{eq.dyn.rashba2} for the particular case of the Fr\"ohlich interaction arising from long-wavelength longitudinal-optical phonons in polar crystals. To this aim we employ Fr\"ohlich's model.\cite{Froehlich1954}

In Fr\"ohlich's model one considers an electron near a parabolic band minimum at the zone center, with effective mass $m^*$, which interacts with a longitudinal optical phonon with frequency $\w$. The variation of the mean-field potential associated with this phonon is:\cite{Sio2019}
  \begin{equation}\label{eq.feg.pot}
  \Delta_{\bq\nu}V = i\left[\frac{e^2}{4\pi\epsilon_0}\frac{4\pi}{\Omega}\frac{\hbar\w}{2}\frac{1}{\k}
  \right]^{1/2}\!\!\frac{\bq\cdot\be_\nu}{\,|\bq|^2}\, e^{i\bq\cdot\br},
  \end{equation}
where $\be_\nu$ is the dimensionless and normalized polarization vector of the mode, $e$ is the electron charge, $\epsilon_0$ the permittivity of vacuum, and $1/\kappa = 1/\ve_\infty - 1/\ve_0$ with $\ve_\infty$/$\ve_0$ the high-frequency/static relative dielectric constant, respectively; $\Omega$ is the unit cell volume.  In the Fr\"ohlich model the degenerate spinor states are written as free electron states, since the phonon-induced potential is structureless on the scale of the crystal unit cell. After a unitary rotation, these states are given by:
  \begin{equation}\label{eq.feg.spinors}
  \psi_{\bk s}(\br) = (N \Omega)^{-1/2} e^{i\bk\cdot \br} \chi_s~,
  \end{equation}
where $\chi_1 = [1\,\, 0]^{\rm T}$ and $\chi_2 = [0\,\, 1]^{\rm T}$.
By replacing Eqs.~(\ref{eq.feg.pot})-(\ref{eq.feg.spinors}) inside Eq.~(\ref{eq.g}) we obtain:
  \begin{equation} \label{eq.g.feg.3}
  g^{s' s}_\nu(\bk+\bq, \bk) =  \gamma \frac{\bq\cdot \be_\nu}{\,|\bq|^2}\,\bk\times \bq
  \cdot \bs_{s's} \qquad \mbox{with }\,\, \gamma = \left[\frac{e^2}{4\pi\epsilon_0}\frac{4\pi}{\Omega}
  \frac{\hbar\w}{2}\frac{1}{\k} \right]^{1/2} \!\!\!\frac{\hbar^2}{4m_e^2c^2}~.
  \end{equation}
Now we consider the case of $\bq$ along the direction of $\be_\nu$. This choice provides an upper bound to the magnitude of the splitting. By setting the reference frame such that $\be_\nu$ points along the $z$ axis, in the limit $\bq\rightarrow 0$ Eq.~\eqref{eq.g.feg.3} becomes:
  \begin{equation} \label{eq.g.feg.3b}
  \lim_{\bq\rightarrow 0}\,g^{s' s}_\nu(\bk, \bk+\bq) = i \gamma
  \begin{pmatrix}
    0 & k_x-ik_y \\ -k_x-ik_y & 0
  \end{pmatrix},
  \end{equation}
The dynamic Rashba-Dresselhaus coupling constant $C_{n\nu}(\bm{\hat{k}})$ of Eq.~\eqref{eq.C.rashba} becomes:
  \begin{equation}
  C_{n\nu}(\bm{\hat{k}}) = \gamma |\cos \theta|,
  \end{equation}
where $\theta$ is the angle between $\bk$ and the $z$ axis. The maximum value is reached for $\theta = n \pi$, with $n$ an integer. In this case the Rashba energy of Eq.~\eqref{eq.dyn.rashba2} takes the form:
  \begin{equation}\label{eq.ER}
  E_{\rm R} = \frac{\pi}{4} \,\alpha^4\,
   \frac{1}{\k} \frac{m^*}{m_e}
  \frac{u^2 \hbar\w}{\Omega/a_0^3}~,
  \end{equation}
where $\alpha\simeq 1/137$ is the fine structure constant. %

The presence of the fine structure constant to the fourth power in Eq.~\eqref{eq.ER} makes the Rashba energy associated with longitudinal optical phonons negligible. For example, in the case of methylammonium lead iodide, using $m^* = 0.1\,m_e$,\cite{Miyata2015} $\varepsilon_\infty = 6.5 \epsilon_0$,\cite{Hirasawa1994} $\varepsilon_0 = 30.5\epsilon_0$,\cite{Poglitsch1987} $\Omega = 6407\,a_0^3$,\cite{Whitfield2016} we obtain $E_{\rm R} \simeq  10^{-13}\,u^2\hbar\w$. Even for highly-energetic coherent phonons with $u^2\hbar\w \sim 100$~meV per unit cell, the Rashba energy would be smaller than $10^{-11}$~meV.

\section{Dynamic Rashba-Dresselhaus Hamiltonian from k$\cdot$p perturbation theory} \label{note-9}

We analyze the symmetry of the Rashba-Dresselhaus Hamiltonian for coherent states. The effective Hamiltonian is given by Eqs.~\eqref{eq.expt.values.4}-\eqref{eq.delta}, which we rewrite in compact form:
  \begin{equation}\label{eq.symm1}
  H_{{\rm tot},s s'} = (\ve_{n\bk} + u^2 N_p \hbar\w_{\bq\nu})
     \d_{ss'} + \d_{\bq,0}\,2 u\, g^{ss'}_{nn\nu}(\bk,\bk)~.
  \end{equation}
The spin splitting arises from the $g^{ss'}_{nn\nu}(\bk,\bk)$ term on the right hand side, therefore we focus on this term. From Eq.~\eqref{eq.g} we have:
  \begin{equation} \label{eq.symm2}
  g_{nn\nu}^{s' s}  (\bk, \bk) = \prefac \< \psi_{n\bk s'}| \bs \cdot \nabla
  (\Delta_{0\nu} V) \times \bp\, | \psi_{n\bk s}\>~.
  \end{equation}
After expressing the spinors as Bloch states, $\psi_{n\bk s} = N^{-1/2} u_{n\bk s}\exp(i\bk\cdot\br)$,
we have:
  \begin{equation} \label{eq.symm3}
  g_{nn\nu}^{s' s}  (\bk, \bk) = \prefac \< u_{n\bk s'}| \bs \cdot \nabla
  (\Delta_{0\nu} V) \times (\bp+\hbar \bk) | u_{n\bk s}\>_{\rm uc}~,
  \end{equation}
where the subscript `uc' indicates that the integral is over one crystal unit cell.  We now express the states $u_{n\bk s}$ for small $\bk$ using $\bk\cdot\bp$ perturbation theory.  The total $\bk$-projected Hamiltonian without electron-phonon interactions is:
  \begin{equation} \label{eq.kdotp}
   \hH_{\bk} = \hH_{\bk=0} + \hH_{\bk\cdot\bp} \qquad \mbox{with} \qquad
      \hH_{\bk\cdot\bp} = \frac{\hbar^2 |\bk|^2}{2m_e} + \frac{\hbar}{m_e} \bk \cdot \left(
     \bp + \frac{\hbar}{4m_e c^2}\bs\times \nabla V\right)~.
  \end{equation}
Since $u_{n\bk s}$ and $u_{n\bk s'}$ are degenerate as a result of parity and time-reversal, to determine their mixing at zeroth order we need degenerate perturbation theory.  To this aim we must diagonalize the following matrix:
  \begin{equation}
  \< u_{n 0 s'}| \hH_{\bk\cdot\bp} | u_{n 0 s}\>_{\rm uc}
      =  \bk \cdot \frac{\hbar}{m_e}\< u_{n 0 s'}|   \left(
     \bp + \frac{\hbar}{4m_e c^2}\bs\times \nabla V\right) | u_{n 0 s}\>_{\rm uc}~.
  \end{equation}
Using Eq.~\eqref{eq.kramers} and the fact that the self-consistent potential $V(\br)$ is even
under parity and real-valued, after some algebra we obtain:
  \begin{equation}
    \< u_{n 0 1}| \hH_{\bk\cdot\bp} | u_{n 0 1}\>_{\rm uc}
    = \< u_{n 0 2}| \hH_{\bk\cdot\bp} | u_{n 0 2}\>_{\rm uc}
     \in {\rm Re}~, \quad
    \< u_{n 0 1}| \hH_{\bk\cdot\bp} | u_{n 0 2}\>_{\rm uc}
    = \< u_{n 0 2}| \hH_{\bk\cdot\bp} | u_{n 0 1}\>_{\rm uc} = 0~.
  \end{equation}
Therefore $\hH_{\bk\cdot\bp}$ does not mix the states $u_{n 0 1}$ and $u_{n 0 2}$, and the degeneracy is not lifted, as expected.  To obtain the correction to the wavefunctions to first order we use the standard sum-over-states, except that we must now exclude the degenerate subspace:
  \begin{equation}
    u_{n \bk s} = u_{n 0 s} + \sum_{m\ne n,s'} \frac{\<u_{m 0 s'}|\hH_{\bk\cdot\bp}| u_{n 0 s}\>_{\rm uc}}
        {\ve_{n 0 s}-\ve_{m 0 s'}}u_{m 0 s'}~.
  \end{equation}
Using Eq.~\eqref{eq.kdotp} inside this expression, and replacing the result inside Eq.~\eqref{eq.symm3}, we obtain:
  \begin{eqnarray}
  g_{nn\nu}^{s' s}  (\bk, \bk) &=& \prefac \left[\< u_{n 0 s'} | \bs \cdot \nabla
  (\Delta_{0\nu} V) \times (\bp+\hbar \bk) | u_{n 0 s} \>_{\rm uc} \right. \nonumber \\
     &+& \left. \hbar\bk \cdot \!\!\!\!\!\sum_{m\ne n,s''} \frac{
    A_{mn s''}^{s's}(\bk) +A_{mn s''}^{s s',*}(\bk) } {\ve_{n 0 s}-\ve_{m 0 s''}}\right]+{O}(k^2)~,
  \end{eqnarray}
having defined:
  \begin{equation}
   A_{mn s''}^{s's}(\bk) =\frac{1}{ m_e}
    \< u_{n 0 s'} | \bs \cdot \nabla (\Delta_{0\nu} V) \times (\bp+\hbar \bk) |u_{m 0 s''} \>_{\rm uc}
   \<u_{m 0 s''}|    \left( \bp + \frac{\hbar}{4m_e c^2}\bs\times \nabla V\right) | u_{n 0 s}\>_{\rm uc}~.
  \end{equation}
Now we can use Eq.~\eqref{eq.g.odd.zero} to find that $g_{nn\nu}^{s s'}(0, 0) = 0$ for odd-parity phonons, therefore the last expression simplifies into:
  \begin{equation}
  g_{nn\nu}^{s' s}  (\bk, \bk) = \bk \cdot \prefact \left[ \< u_{n 0 s'} | \bs \times \nabla
  (\Delta_{0\nu} V) | u_{n 0 s} \>_{\rm uc}
     +\sum_{m\ne n,s''} \frac{
    A_{mn s''}^{s's}(0) +A_{mn s''}^{s s',*}(0) } {\ve_{n 0 s}-\ve_{m 0 s''}}\right]+{O}(k^2)~.
  \end{equation}
In the following we ignore the second term since the denominators are of the order of interband transition energies, hence this term only provides a small correction to the first term.

Using the Levi-Civita symbol $\epsilon_{\a\b\g}$ and the Einstein convention, the first term can be written as:
 \begin{equation}
  g_{nn\nu}^{s' s}(0, 0) = \epsilon_{\a\b\g} k_\a  \prefact
       \< u_{n 0 s'}|  \s_\b \D_\g (\Delta_{0\nu} V)
        | u_{n 0 s}\>_{\rm uc}~.
\end{equation}
This is a traceless $2\times 2$ matrix in the spinor labels $s,s'$, therefore it can be expressed in terms of Pauli matrices as:
  \begin{equation}
  2 g^{s's} = (g^{12}+g^{21})\s_x + i(g^{12}-g^{21})\s_y + (g^{11}-g^{22})\s_z~,
  \end{equation}
having omitted obvious indices for clarity. Using this decomposition, and by denoting as $\bg$ the matrix whose elements are $g^{s's}$, we can write:
  \begin{eqnarray}
  \bg &=&
           \prefact\frac{1}{2}
      \epsilon_{\a\b\g} k_\a \Big\{  \nonumber \\
   &&  \phantom{i} \left[ \< u_{n 0 1}|  \s_\b \D_\g (\Delta_{0\nu} V)  | u_{n 0 2}\>_{\rm uc}
     +\< u_{n 0 2}|  \s_\b \D_\g (\Delta_{0\nu} V)  | u_{n 0 1}\>_{\rm uc}  \right]\s_x \nonumber \\
   &+& i\left[ \< u_{n 0 1}|  \s_\b \D_\g (\Delta_{0\nu} V)  | u_{n 0 2}\>_{\rm uc}
     -\< u_{n 0 2}|  \s_\b \D_\g (\Delta_{0\nu} V)  | u_{n 0 1}\>_{\rm uc}  \right]\s_y \nonumber \\
   &+& \phantom{i} \left[ \< u_{n 0 1}|  \s_\b \D_\g (\Delta_{0\nu} V)  | u_{n 0 1}\>_{\rm uc}
     -\< u_{n 0 2}|  \s_\b \D_\g (\Delta_{0\nu} V)  | u_{n 0 2}\>_{\rm uc}  \right]\s_z
      \Big\}~.
  \end{eqnarray}
Now we can use the algebra of Pauli matrices to recast $\bg$ in the desired form:
  \begin{equation}
  \bg =  k_\a G_{\a\b }\,\s_\b~,
  \end{equation}
where the $3\times 3$ matrix $G_{\a\b}$ is given by:
  \begin{eqnarray}
    2\frac{4m_e^2 c^2}{\hbar^2}G_{\a\b }
   &=& \< u_{n 0 1}| [ \bs\times\nabla]_\a (
          \Delta_{0\nu} V)  | u_{n 0 1}\>_{\rm uc}\,\d_{\b 3} \nonumber \\
   &+&    \< u_{n 0 1}|[ \bs\times\nabla]_\a
       ( \Delta_{0\nu} V ) | u_{n 0 2}\>_{\rm uc}\, (\d_{\b 1}+i\d_{\b 2} )
      \nonumber \\
    &+&  \< u_{n 0 2}|[ \bs\times\nabla]_\a
       (\Delta_{0\nu} V )
        | u_{n 0 1}\>_{\rm uc} \,(\d_{\b 1}-i\d_{\b 2}) \nonumber\\
    &-& \< u_{n 0 2}|[ \bs\times\nabla]_\a (\Delta_{0\nu} V) |
            u_{n 0 2}\>_{\rm uc} \,  \d_{\b 3}~. \label{eq.symm6}
  \end{eqnarray}
This matrix is real-valued. The nonzero elements of this matrix can be identified by analyzing the symmetry of the spinors $u_{n 0 1}$ and $u_{n 0 2}$ and of the vibrational eigenmode $e_{\k\a,\nu}(0)$.  { To this aim, we can decompose the matrix $G$ into three terms: {isotropic ($G^{\rm I}$), traceless} symmetric ($G^{\rm S}$), and antisymmetric ($G^{\rm A}$).  With this decomposition the Hamiltonian splits in three components,
$H = H^{\rm I}+H^{\rm S}+H^{\rm A}$:\\
  \begin{eqnarray}
  H^{\rm I} &=& G^{\rm I}_{11}\,k_x \s_x + G^{\rm I}_{22}\,k_y \s_y + G^{\rm I}_{33}\,k_z \s_z, \\
  H^{\rm A} &=& G^{\rm A}_{12}\,(k_x \s_y-k_y \s_x) + G^{\rm A}_{13}\,(k_x \s_z-k_z \s_x)
    + G^{\rm A}_{23}\,(k_y\s_z-k_z\s_y), \\
  H^{\rm S} &=& G^{\rm S}_{12}\,(k_x \s_y+k_y \s_x) + G^{\rm S}_{13}\,(k_x \s_z+k_z \s_x) +
    G^{\rm S}_{23}\,(k_y\s_z+k_z\s_y).
  \end{eqnarray}
{The Hamiltonian $H^{\rm I}$ has the same form found in Weyl semimetals.} $H^{\rm A}$ is a three-dimensional Rashba Hamiltonian. $H^{\rm S}$ represents a linear Dresselhaus Hamiltonian. Therefore the spin pattern induced by a coherent phonon is dictated by the symmetry of the $G$ matrix.


  \begin{table}[b!]
    \caption{Selection rules for the dynamic Rashba-Dresselhaus effect in the
$D_{\rm 2h}$ point group. The irreducible representation $\Gamma_{\a\nu}$ of the
operator $[\bs \times \nabla]_\a (\Delta_{0\nu} V)$ directly determines which matrix
elements $\<u_{n0s'}| [\bs \times \nabla]_\a (\Delta_{0\nu} V)] | u_{n0s} \>$ are
nonzero. (a)
Coupling constants of the $D_{\rm 2h}$
    point group, from Table~19 of Ref.~\onlinecite{Koster1963}. (b) The irreducible
representation of the phonon mode determines the symmetry of the operator
$\Gamma_{\a\nu}$ and thereby which couplings coefficients, i.e. matrix elements of $G$,
are nonzero.\vspace{10pt}}
    \label{tab-s1}
    \centering
  \begin{tabular}{*{7}{c}}
    \multicolumn{1}{l}{(a)} & & & \\
     \hline
     \hline
    \multicolumn{1}{l}{} & \multicolumn{2}{c}{$\Gamma_{\a\nu} = B_{\rm1g}$} &
\multicolumn{2}{c}{$\Gamma_{\a\nu} = B_{\rm2g}$} & \multicolumn{2}{c}{$\Gamma_{\a\nu} =
B_{\rm3g}$} \\
    & $|u_{n01}\>$ & $|u_{n02}\>$ & $|u_{n01}\>$ & $|u_{n02}\>$ & $|u_{n01}\>$ &
$|u_{n02}\>$
    \\ \hline
    $\<u_{n01}|$ & $i$ &  0   &  0 & 1   & 0 & $i$ \\
    $\<u_{n02}|$ & 0 & $-i$   & $-1$ & 0   & $i$ & 0 \\
     \hline
     \hline
\vspace{10pt}
  \end{tabular}

  \begin{tabular}{*{5}{c}}
    \multicolumn{1}{l}{(b)} \\
    \hline
    \hline
    & \multicolumn{3}{c}{$\Gamma_{\a\nu}$} \\
    \cline{2-4}
    \phantom{xx}Mode\phantom{xx}&$\a = 1$    & $\a = 2$    & $\a = 3$    & couplings \\
    \hline
    $A_{\rm u}$ & $B_{\rm3g}$ & $B_{\rm2g}$ & $B_{\rm1g}$ & \phantom{xx}$k_x \s_x$, $k_y \s_y$, $k_z
\s_z$\phantom{xx} \\
    $B_{\rm1u}$ & $B_{\rm2g}$ & $B_{\rm3g}$ & $A_{\rm g}$ & $k_x \s_y$, $k_y \s_x$ \\
    $B_{\rm2u}$ & $B_{\rm1g}$ & $A_{\rm g}$ & $B_{\rm3g}$ & $k_x \s_z$, $k_z \s_x$\\
    $B_{\rm3u}$ & $A_{\rm g}$ & $B_{\rm1g}$ & $B_{\rm2g}$ & $k_y \s_z$, $k_z \s_y$\\
    \hline
    \hline
  \end{tabular}
  \end{table}

\clearpage\newpage

\begin{figure*}
\centering
\includegraphics[width=\textwidth]{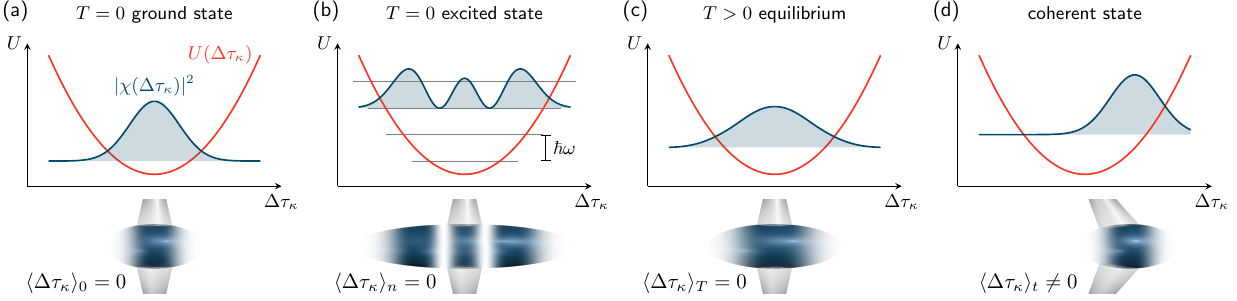}
\vspace{1pt}\caption{
Schematic representations of the square modulus of the quantum nuclear wavefunction
$\lvert\chi(\Delta \tau_\kappa)\rvert^2$ (blue) in a parabolic potential well $U(\Delta \tau_\kappa)$
(red). (a) The probability is inversion symmetric in the ground state. (b)
The probability is inversion symmetric in any excited state at $T = 0$.
(c) The probability is inversion symmetric in thermodynamic
equilibrium at $T > 0$. (d) A quantum nuclear wavefunction breaking inversion symmetry
can be obtained by exciting a coherent phonon.
}
\label{fig-s1}
\end{figure*}

  \begin{figure}
  \centering
  \includegraphics[width=0.7\textwidth]{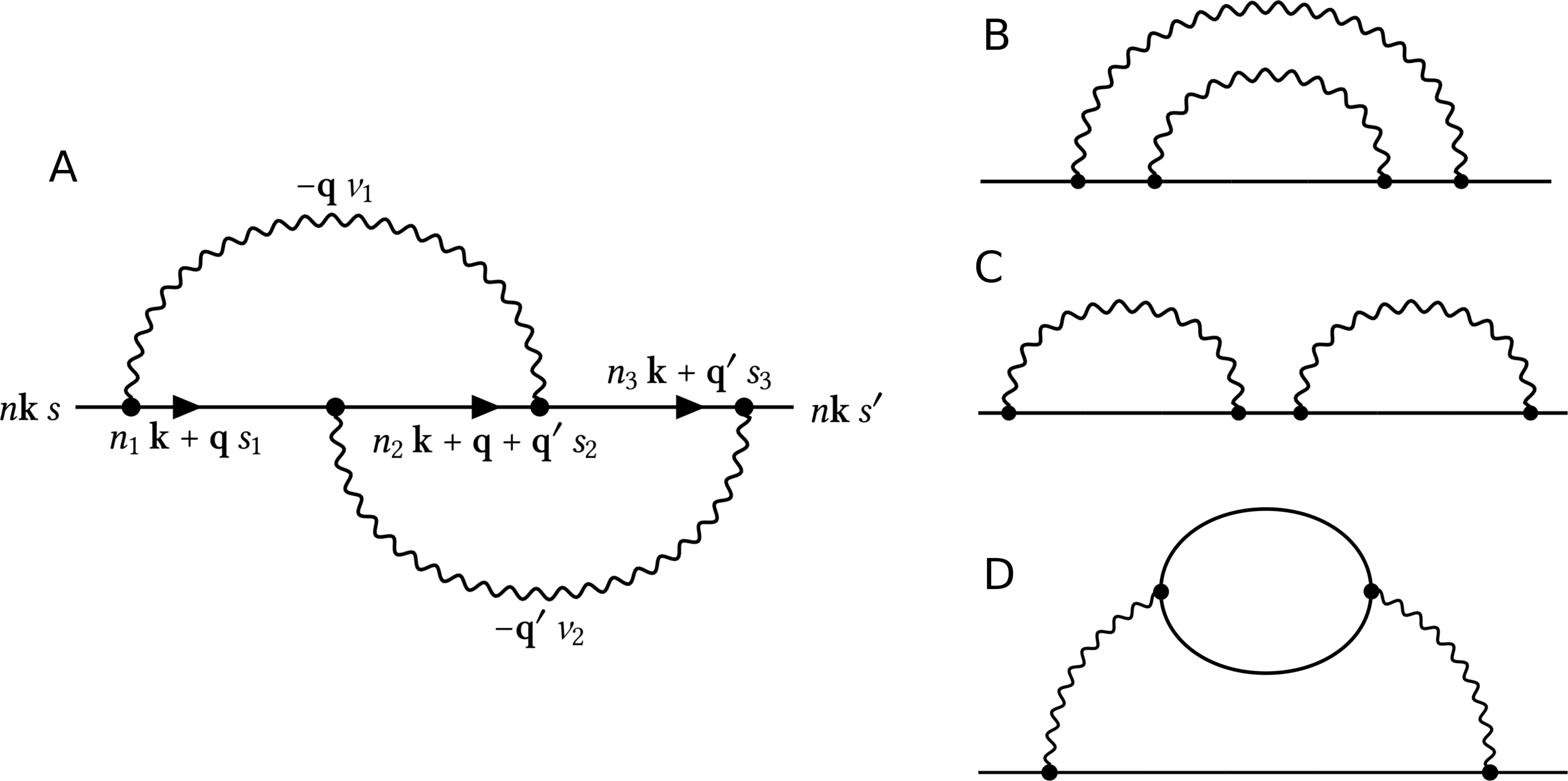}
  \vspace{5pt}\caption{\label{fig-s2}
  Inequivalent Feynman diagrams for the virtual scattering processes contributing to the electron energy renormalization to fourth order in the spin-phonon interaction. Electrons and phonons are denoted by straight and wiggly lines, respectively. The dots represent the spin-phonon matrix elements. The band, spin, and momentum labels indicated in A were used to obtain Eq.~\eqref{eq.4th}.
  }
  \end{figure}

  \begin{figure}[h]
  \centering
  \includegraphics[width=0.7\textwidth]{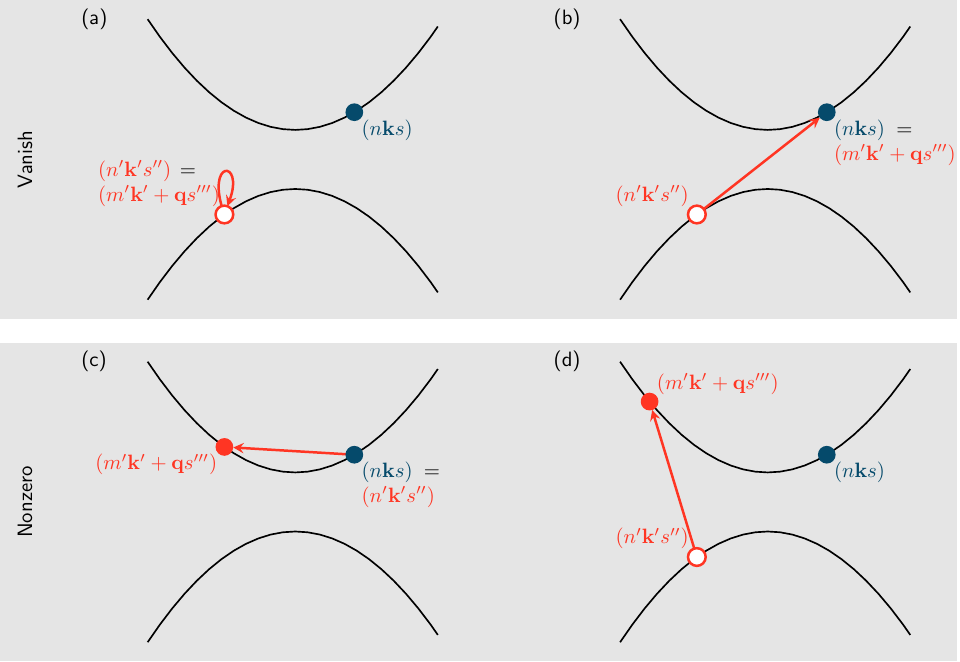}
  \caption{\label{fig-s3}
  Virtual processes contributing to the energy renormalization of the state $|n\bk s\>$.  Process~(a) vanishes for an extended crystal due to the conservation of crystal momentum, while (b) vanishes due to Pauli blocking. Process (c) is an allowed virtual transition, and (d) describes a fluctuation of the Fermi vacuum. The parabolas represent idealized energy bands, and filled/empty disks denote electrons/holes, respectively.
  }
  \end{figure}

  \begin{figure}
  \centering
  \includegraphics{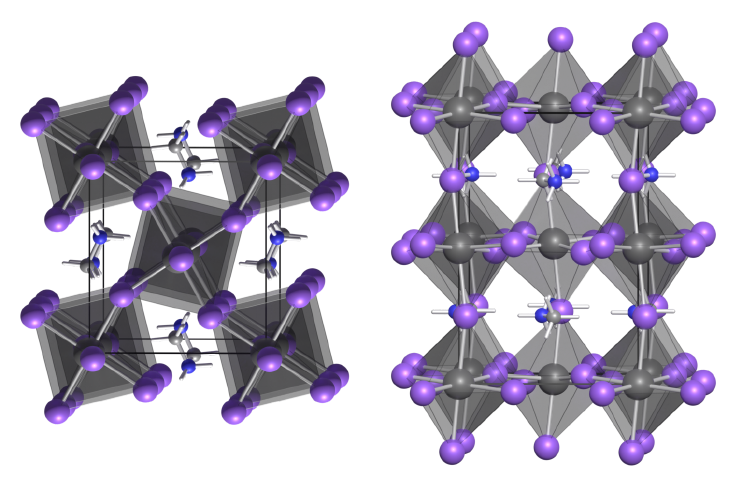}
  \caption{Schematic ball-stick models of orthorhombic $Pnma$ MAPI, consisting of Pb (darkgray), I (purple), C (lightgray), N (blue), and H (white).
  \label{fig-s4}
  }
  \end{figure}


%

\end{document}